\documentclass[a4paper,11pt]{article}

\usepackage{jcappub}
\usepackage{color}
\usepackage{float}
\usepackage{amssymb}
\usepackage{amsmath}
\usepackage{slashed}
\usepackage{mathtools}
\usepackage{xcolor}

\newcommand{\Neff}{N_\mathrm{eff}}
\newcommand{\DNeff}{\Delta N_\mathrm{eff}}
\newcommand{\DNeffCMB}{\Delta N_\mathrm{eff}^\mathrm{CMB}}
\newcommand{\eV}{\mathrm{eV}}

\newcommand{\MeV}{\mathrm{MeV}}

\newcommand{\ML}[1]{#1}
\newcommand{\MLc}[1]{}

\title{Novel cosmological bounds on thermally-produced axion-like particles}

\author[a,b]{Luca Caloni}
\author[b]{\!,\, Martina Gerbino}
\author[b]{\!,\\ Massimiliano Lattanzi}
\author[c,d]{\!,\, Luca Visinelli}
\affiliation[a]{Dipartimento di Fisica e Scienze della Terra, Universit\`a degli Studi di Ferrara,\\ Via Giuseppe Saragat 1, I-44122 Ferrara, Italy}
\affiliation[b]{Istituto Nazionale di Fisica Nucleare, Sezione di Ferrara,\\ Via Giuseppe Saragat 1, I-44122 Ferrara, Italy}
\affiliation[c]{Tsung-Dao Lee Institute (TDLI),\\
520 Shengrong Road, 201210 Shanghai, P.\ R.\ China}
\affiliation[d]{School of Physics and Astronomy, Shanghai Jiao Tong University,\\
800 Dongchuan Road, 200240 Shanghai, P.\ R.\ China}

\emailAdd{luca.caloni@unife.it}
\emailAdd{gerbinom@fe.infn.it}
\emailAdd{lattanzi@fe.infn.it}
\emailAdd{luca.visinelli@sjtu.edu.cn}

\date{today}

\abstract{We constrain the coupling of thermally-produced axion-like particles (here axions) with photons and gluons, using data from the cosmic microwave background (CMB) spectra and baryon acoustic oscillations. The axion possesses an explicit soft breaking mass term and it is produced thermally in the early Universe from either axion-photon or axion-gluon processes, accounting for the recent progresses in the field. We derive the most stringent bounds on the axion-gluon coupling to date on the mass range considered $10^{-4} \lesssim m_a/{\rm eV} \lesssim 100$, superseding the current bounds from SN1987A. The bounds on the axion-photon coupling are competitive with the results from the CAST collaboration for the axion mass $m_a \gtrsim 3\,$eV. We comment on the forecast reaches that will be available given the sensitivity of future CMB-S4 missions.}

\begin{document}
\maketitle
\flushbottom

\section{Introduction}
\label{sec:intro}

The Peccei-Quinn (PQ) mechanism~\cite{Peccei:1977hh, Peccei:1977ur}, one of the most appealing solutions to the strong-CP problem of the Standard Model (SM), introduces a new Abelian U(1)$_{\rm PQ}$ symmetry which is anomalous under gauged interactions and spontaneously broken below an energy scale $f_a$. The QCD axion is the pseudo Nambu-Goldstone boson of this broken symmetry~\cite{Weinberg:1977ma, Wilczek:1977pj, Goldman:1977en}. The axion mass and all couplings to matter and gauge bosons are inversely proportional to $f_a$, which lies well above the electroweak scale in order to avoid current experimental constraints~\cite{Graham:2015ouw, Irastorza:2018dyq, Sikivie:2020zpn}. If this picture is correct, axions would have been produced in the early Universe both through thermal and non-thermal processes. Mechanisms that produce a non-thermal population include vacuum realignment~\cite{Abbott:1982af, Dine:1982ah, Preskill:1982cy} and the decay of topological defects~\cite{Vilenkin:1982ks, Huang:1985tt, Davis:1986xc}, both leading to a present cold axion density that could potentially address the origin of dark matter (DM). Thermal axions~\cite{Turner:1986tb} are produced through various interaction with both SM and other speculative particles, representing a hot DM component and forming a background of cosmic axions.

Along with the QCD axions, other light particles with similar properties could arise from various extensions of the SM, such as the spontaneous breaking of additional global symmetries~\cite{McDonald:1993ex, Burgess:2000yq, He:2009yd, Cacciapaglia:2019bqz, Lin:2022xbu}, Goldstone modes from string theory compactification~\cite{Svrcek:2006yi, Arvanitaki:2009fg}, or accidental symmetries~\cite{Harigaya:2013vja, DiLuzio:2017tjx}. Likewise, these axion-like particles would contribute to the present dark matter budget~\cite{Arias:2012az, Visinelli:2017imh}. In particular, models of string theory generically predict the production of axions from the decay of moduli fields, which would appear at present time as a homogeneous axion background. This axion component can lead to various electromagnetic and gravitational signals, such as the axion-photon conversion in intergalactic media or the appearance of superradiance around spinning black holes. In the following, we refer to an axion-like particle simply as an ``axion''.

The assessment of the present background of hot axions depends on the specific production mechanism and on the temperature $T_d$ at which the axion decouples from the primordial plasma. If coupled with gluons, axions would scatter off the quark-gluon plasma well above the QCD phase transition~\cite{Masso:2002np, Graf:2010tv, Salvio:2013iaa}, while they would be produced through hadronic scatterings at temperatures below the QCD phase transition~\cite{Berezhiani:1992rk, Chang:1993gm, Hannestad:2005df, DEramo:2014urw, Kawasaki:2015ofa, Ferreira:2020bpb}. The contribution from other production mechanisms such as the Primakoff process depends on the relative strength of the coupling of the axion with SM particles such as the electron and the photon, and could either overcome the production off the quark-gluon plasma and be the main source of production or be subdominant. For a list of these processes, see Ref.~\cite{DEramo:2018vss} for the production through lepton channels, Refs.~\cite{Ferreira:2018vjj, Arias-Aragon:2020qtn} for the production from scattering off heavy quarks, and Ref.~\cite{Dror:2021nyr} for the production relying on photon coupling.

Cosmology has proven to be an extraordinary tool to constrain the parameter space of axions and other light particles, because of the sensitivity of some cosmological probes such as cosmic microwave background (CMB) and matter power spectra at intermediate and small angular scales on the amount of dark radiation in excess to what is expressed by SM neutrinos~\cite{Bashinsky:2003tk, Hou:2011ec}. One key parameter to address the presence of new light species through cosmology is the effective number of relativistic species $N_{\rm eff}$ which incorporates, along with neutrinos, any other particle that is relativistic at recombination (see e.g.\ Ref.~\cite{Weinberg:2013kea}). Within SM, the effective number of neutrinos is expected to be $N_{\rm eff}^{\rm SM} = 3.046$ once non-instantaneous decoupling, neutrino oscillations and radiative corrections are accounted for~\cite{Mangano:2001iu}, but see also Refs.~\cite{deSalas:2016ztq, Akita:2020szl, Bennett:2020zkv} for most updated and independent analyses. Any relativistic species other than neutrinos such as light sterile neutrinos or thermal axions would be a dark radiation component that potentially adds up to $N_{\rm eff}$. This excess is parametrized by the deviation from the value expected from SM as $\Delta N_{\rm eff} \equiv N_{\rm eff} - N_{\rm eff}^{\rm SM}$.

The effective number of relativistic species $N_{\rm eff}$ is degenerate with various other observables in the CMB, so that a sharp change in $N_{\rm eff}$ would impact on the reconstruction of other cosmological quantities as inferred from observations. The degeneracy is partially removed when low-redshift baryon acoustic oscillations (BAO) data are included in the analysis along with CMB results. The most updated constraints on $\Delta N_{\rm eff}$ from the {\it Planck} 2018 collaboration sets $N_{\rm eff} = 2.99\pm0.17$ at 68\% confidence level (CL), using a combination of TT,TE,EE+lowE data~\cite{Planck:2018vyg} plus observations from BAO. Future surveys will be extremely sensitive to the deviation of $N_{\rm eff}$ from its SM value, with a forecast constraint $\Delta N_{\rm eff} \lesssim 0.06$ at 95\% CL~\cite{CMB-S4:2016ple, Abazajian:2019eic}.

Given the sensitivity of the cosmological data, it is possible to bound the properties of Goldstone bosons from their contribution to $N_{\rm eff}$. In fact, for an exact broken symmetry the Goldstone boson would be massless, although a light mass is expected by the spontaneous breaking of an approximate symmetry. Various models emerge within the spontaneous breaking of new hypothetical symmetries such as axions, majorons, and familons. Restricting the analysis to the QCD axion, a recent combination of measurements from CMB, BAO, and supernovae data bounds the mass of the QCD axion $m_a \lesssim 7.46\,$eV at 95\% CL, assuming a predominant production from gluon scattering, and $m_a \lesssim 0.91\,$eV at 95\% CL, assuming the axion is produced from scattering off pions at decoupling temperatures $T_d \lesssim 62\,$MeV~\cite{Giare:2020vzo}, see also Refs.~\cite{Hannestad:2003ye, Hannestad:2007dd, Melchiorri:2007cd, Archidiacono:2013cha, Giusarma:2014zza, DiValentino:2014zna, DiValentino:2015wba} for earlier work on the subject.

In this work, we focus on axions which are produced thermally in the early Universe from processes involving the couplings to gluons and photons. In the case of the QCD axion, the contribution from the axion-gluon coupling is always dominant in establishing a thermal population of axions, see e.g.\ Ref.~\cite{Salvio:2013iaa}. More generally, the couplings of an axion with SM particles can vary over wider ranges. Motivated by a phenomenological approach, we treat the two production mechanisms separately and we assume that only one of them dominantly produces the majority of the hot axions. We focus here on the combination of {\it Planck} 2018 and BAO data as representative of the constraining power of current cosmological data while restricting to linear and mildly non-linear scales and probes of the background expansion. We do not consider additional data from, e.g., measurements of small-scale CMB fluctuations from ACT~\cite{ACT:2020gnv,ACT:2020frw} and SPT~\cite{SPT-3G:2021wgf,SPT-3G:2021eoc}, broad-band shape of the matter power spectrum and weak lensing~\cite{DES:2022qpf,DES:2021wwk,BOSS:2016wmc,KiDS:2020suj,KiDS:2020ghu}. Their inclusion, together with the detailed analysis of the impact of non-linear modelling, mild inter-dataset disagreements and their cross-covariances (see e.g.\ Refs.~\cite{Sgier:2021bzf,ACT:2020gnv}), goes beyond the scope of the current manuscript and will be addressed in future work. 
We will however briefly comment on prospects for constraints on the axion couplings including small-scale CMB data.

The paper is organized as follows. In Sec.~\ref{sec:cosmology} we introduce an effective axion Lagrangian that includes the couplings with the photon and the gluon fields, and we discuss the relation between the axion abundance and $\Delta N_{\rm eff}$. In Sec.~\ref{sec:decoupling} we comment on the production of axions in the early Universe. In Sec.~\ref{sec:analysis} we present the datasets used in the Monte Carlo analysis that we perform to derive the bounds on the thermally-produced axions. Results for the runs are given in Sec.~\ref{sec:results} and conclusions are drawn in Sec.~\ref{sec:conclusions}. We work in units with $\hbar = c = 1$.

\section{Cosmology of thermal axions}
\label{sec:cosmology}

In this section we outline the mechanism to thermally produce the axion in the early Universe. \ML{Before doing so, however, we give a glimpse of the main results of the paper.} 
In this study, we obtain a limit on the temperature $T_d$ at which the axion ceases to be chemically coupled to the primordial plasma as a function of the effective axion mass $m_a$.
\ML{This limit is essentially a bound on the axion abundance. For smaller axion masses, it is dominated by observational constraints on the axion contribution to the radiation energy density in the early Universe, or equivalently on the deviation $\Delta N_{\rm eff}$ of the effective number of relativistic species from its SM value. In the opposite regime of larger axion masses, limits on $T_d$ are instead obtained from constraints on the axion contribution to the energy density of nonrelativistic matter.}

\ML{Since the parameter space explored in our analysis probes a wide range of values of the axion mass (more precisely, of the axion mass-to-temperature ratio $m_a/T$), we find useful for the discussion to distinguish between the value of $\Delta N_{\rm eff}$ which is obtained in the limit of vanishing mass, from the actual contribution of axions to the total energy density at the time of recombination $t_\mathrm{rec}$. We will use the notation $\Delta N_{\rm eff}$, without any specification, for the former quantity, while we denote the latter quantity with $\Delta N_{\rm eff}^{\rm CMB}$.}

More specifically, $\Delta N_{\rm eff}$ can be obtained analytically from the definition in Eq.~\eqref{eq:Delta-Neff} and is essentially a proxy for the axion temperature relative to the photons, $T_a/T_\gamma$. Instead, $\Delta N_{\rm eff}^{\rm CMB}$ accounts for the finite effects of a non-zero mass at recombination. These two quantities are related by Eq.~\eqref{eq:DNeff-CMB}. The usefulness of $\Delta N_{\rm eff}^{\rm CMB}$ is that CMB constraints on the energy density of relativistic species apply more directly to this quantity, rather than to $\DNeff$. This is somehow a simplification; it is by now understood (see e.g.\ Ref.~\cite{Hou:2011ec}) that the constraining power of present CMB experiments on $\Neff$ mainly derives from their ability to measure the ratio between two scales, the sound horizon and the Silk damping length at recombination. These scales in principle depend on the integrated expansion history between reheating and recombination, but they are dominated by the contribution from the larger redshifts in the integral. Thus, even if the effect of relativistic species on CMB observables cannot be completely captured by just looking at their energy density evaluated at recombination time, yet we take $\DNeffCMB$ as a good enough proxy to roughly assess whether a given axion model satisfies CMB constraints on $\Neff$. In practice, $\DNeffCMB$ has the nice properties of i) being equal to $\DNeff$ in the limit $m_a/T_a|_{t=t_\mathrm{rec}} \ll 1$, when the CMB constraints on $\Neff$ should apply at face value, while ii) being $\ll \DNeff$ in the opposite regime $m_a/T_a|_{t=t_\mathrm{rec}} \gg 1$, when these same constraints should not be relevant. We further stress that $\DNeffCMB$ is only used in the qualitative discussion in this section, and, possibly, in later parts of the paper only to gain physical insight on which observational features drive the constraints on axion parameters. In our complete analysis, we always integrate numerically the evolution of background and perturbation quantities for all species, including axions, properly taking into account the effect of the finite axion mass across all the cosmological evolution. In our framework, axion imprints on cosmological observables depend on two quantities: the effective axion mass $m_a$ and the present axion temperature $T_{a0}$, or equivalently the axion decoupling temperature $T_d$, see Eq.~\eqref{eq:Ta}. These two quantities can be effectively traded for any two among $\Delta N_{\rm eff}$, $\Delta N_{\rm eff}^{\rm CMB}$ or the present energy density parameter $\omega_a$ defined in Eq.~\eqref{eq:axion-density} below.

\begin{figure}[ht]
	\centering
	\includegraphics[width=0.7\textwidth]{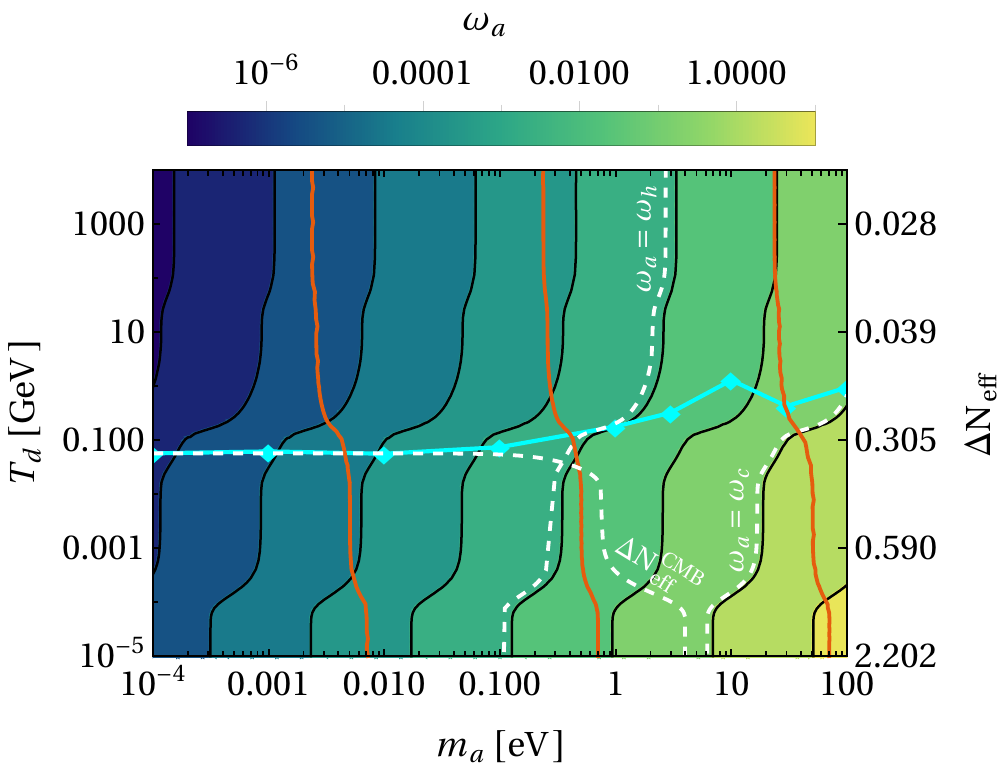}
	\caption{Relic abundance $\omega_a$ of thermal axions as a function of the axion mass $m_a$ and decoupling temperature $T_d$ (or, equivalently, $\Delta N_\mathrm{eff}$). The vertical dashed white lines represent the bounds on the abundance of hot and cold DM as defined in the main text. The horizontal (in the left part of the plot) dashed white line corresponds to $\Delta N_{\rm eff}^{\rm CMB} = 0.376$ (see main text for details). Also shown are the bounds on $T_d$ (solid cyan line) as a function of $m_a$. The red lines represent the pairs of axion parameters with constant values of $\langle p_{a,\mathrm{rec}} \rangle/m_a = \{10^2, 1, 10^{-2}\}$ from left to right.}
	\label{fig:Tdvsma}
\end{figure}

One of the main results of this work is shown in Fig.~\ref{fig:Tdvsma}. The color shades report the relic abundance of thermal axions $\omega_a$ as a function of the {\it effective} axion mass $m_a$ in Eq.~\eqref{eq:effective-mass} (horizontal axis) and decoupling temperature $T_d$ (vertical left axis) or, equivalently, the value of $\Delta N_\mathrm{eff}$ (vertical right axis). The scaling of $\Delta N_\mathrm{eff}$ is not logarithmic as this quantity is related to $T_d$ by a complicated function involving the effective number of entropy degrees of freedom $g_{*s}(T)$ at temperature $T$. The contours range from $10^{-7}$ to the top left corner of the figure to $10$ to the bottom right. The dashed white lines mark the current limits on cold dark matter (CDM), $\omega_a < \omega_c$, where $\omega_c = 0.1202 \pm 0.0014$ at 68\% CL from {\it Planck} 2018 results with TT,TE,EE+lowE data~\cite{Planck:2018vyg}. Also shown is the bound on hot dark matter\footnote{The bound on hot dark matter is obtained here by translating cosmological constraints on the sum of neutrino masses $\Sigma m_\nu$~\cite{Planck:2018vyg} to a constraint on their energy density $\omega_h$.} $\omega_a < \omega_h$, with $\omega_h \approx 2\times 10^{-3}$. Both the hot and cold DM bounds indirectly constrain the value of $\Delta N_{\rm eff}$ as a function of the effective axion mass $m_a$ through Eq.~\eqref{eq:omega_a_DNeff}. We also include in Fig.~\ref{fig:Tdvsma} the information on the energy density of light relics at early times. In order to do so, we show the curve of constant $\DNeffCMB = 0.376$. This value corresponds to the 95\% CL upper limit on $\Delta N_\mathrm{eff}$ that we obtain, \emph{for a massless axion} (for which we recall $\DNeffCMB = \DNeff$), from {\it Planck} 2018 TT,TE,EE+lowE data in combination with BAO measurements. 
We further need to distinguish whether the axion is relativistic at recombination. The solid red lines in the plot serve this purpose, marking the curves for which the average axion momentum at recombination, see Eq.~\eqref{eq:effectivemomentum} below, is $\langle p_{a,\mathrm{rec}} \rangle = \{10^2, 1, 10^{-2}\}\,m_a$ from left to right. The axion is nonrelativistic (ultrarelativistic) in the right (left) corner of the plot, when $\langle p_{a,\mathrm{rec}} \rangle \ll m_a$ ($\langle p_{a,\mathrm{rec}} \rangle \gg m_a$). Finally, we plot the bounds on the decoupling temperature, as a function of the axion mass, that we obtain from the full analysis detailed in Sec.~\ref{sec:results_DNeff}. These are shown as the solid cyan line.

We have now the tools to  understand how the axion parameter space is constrained. In the region in which axions are ultrarelativistic at recombination, the relevant constraints are $\DNeffCMB\simeq\DNeff\lesssim 0.376$ and $\omega_a < \omega_h \simeq 2\times 10^{-3}$. In the region in which axions are nonrelativistic at recombination, the only relevant constraint is that coming from the observed dark matter density: $\omega_a < \omega_c \simeq 0.12$. This intuition is nicely confirmed by looking at the bounds obtained by the full, more accurate analysis exactly accounting for the interplay between the axion mass and temperature. In the leftmost part of the plot ($m_a \lesssim 0.1\,\eV$), where axions are effectively massless and behave as dark radiation, the cyan curve reproduces the constrain on $\DNeff$. In the region $m_a \simeq 1\,\eV$, the axion mass starts to be relevant and axions behave as hot dark matter and the full analysis essentially yields the $\omega_a < \omega_h$ bound. Finally, in the rightmost part of the plot, $m_a \gtrsim 30\,\eV$ axions are cold and indeed we reproduce the cold dark matter bound. We stress that, in the ``warm'' axion region $1\,\eV \lesssim m_a \lesssim 30\,\eV$, it is not possible to obtain a useful bound on $T_d$ from a qualitative analysis. It is of course legit to expect that the actual bound will lie between the hot and cold dark matter bounds, but this is of little utility even for the purpose of getting an order-of-magnitude estimate, since as it is clear from the figure the region between the hot and cold DM curves spans many orders of magnitude in $T_d$. In this region, a complete analysis like the one presented in this paper is the only way to obtain meaningful bounds on axion properties. 
The full analysis also helps to uncover features that might fail intuition.
For example, the plot shows that a truly ``cold'' regime is only reached for relatively small values ($\lesssim 10^{-2}$) of the momentum-to-mass ratio at recombination; even for values of $\langle p_{a,\mathrm{rec}} \rangle / m_a$ of a few $\times 10^{-2}$, the constraints are (possibly much) stronger than the ones that one would naively obtain solely from the requirement $\omega_a < \omega_c$. 
In other words, {\it Planck} measurements of the CMB anisotropies are very sensitive even to a little ``warmness'' of axions (and other light relics in general). This was noted, in the different framework of QCD axions in low reheating scenarios, also in Ref.~\cite{Carenza:2021ebx}.

\subsection{The chiral axion Lagrangian}

We now outline the model chosen. An axion is characterized by its anomalous couplings to the gluon and photon fields (see Refs.~\cite{Marsh:2015xka,DiLuzio:2020wdo} for recent reviews). Here, we work under the assumption that below the energy scale $f_a$, the axion field is described by an effective field theory, so that the phenomenology of the light axion is described by the effective Lagrangian
\begin{equation}
    \label{eq:lagrangian}
	\mathcal{L}_{\rm eff} \supset \frac{1}{2}(\partial^\mu a) (\partial_\mu a) - \frac{1}{2}m_0^2a^2 + \mathcal{L}_{ag} + \mathcal{L}_{a\gamma}\,,
\end{equation}
where $a$ is the axion field and the mass term $m_0$ appearing in the Lagrangian yields to a soft explicit breaking of the shift symmetry. The two Lagrangian terms describing the interaction of the axion with the gluon and the photon fields are respectively
\begin{equation}
    \label{eq:axion-gluon}
    \mathcal{L}_{ag} = \frac{\alpha_s}{8\pi} \frac{C_g}{f_a}\,a G_{\mu\nu}^i \tilde{G}^{\mu\nu, i} \,, \quad \mathcal{L}_{a\gamma} = \frac{1}{4} g_{a\gamma}^0 a F_{\mu\nu}\tilde{F}^{\mu\nu} \, ,
\end{equation}
where $\alpha_s$ denotes the strong force coupling constant, $g_{a\gamma}^0$ is the axion-photon coupling, $F_{\mu\nu}$ and $G_{\mu\nu}^i$ are the photon and gluon field strength tensors, $\tilde{F}_{\mu\nu}$ and $\tilde{G}_{\mu\nu}^i$ are their duals, and the index $i$ runs over the adjoint indices of the SU(3) color gauge group. The axion shift symmetry, which is a continuous symmetry at energies above $f_a$, is broken down to a discrete symmetry by the presence of the axion-gluon coupling $C_g$, which is a parameter that modulates the coupling strength of the axion with the gluon field. Within this model, the QCD axion theory is recovered with the choices $C_g = 1$ and $m_0 = 0$.

The axion-gluon coupling in Eq.~\eqref{eq:axion-gluon} induces an irreducible electric dipole moment (EDM) of the neutron $n$ oscillating with time $t$ via a chiral one-loop process as $d_n = g_d a_0 \cos(m_a t)$, where $a_0$ is the local value of the axion field at the position of the nucleus and $g_d$ is the axion-EDM coupling. The Lagrangian term responsible for the neutron EDM can be expressed in terms of a Lorentz invariant operator as~\cite{Crewther:1979pi, Pospelov:1999ha}
\begin{eqnarray}
    \mathcal{L} &\supset& -\frac{i}{2} g_d a \bar{n}\gamma_5\sigma_{\mu\nu}n F^{\mu\nu}\,,    \label{eq:axion_nucleon_coupling}\\
    g_d &=& \frac{C_{an\gamma}}{m_n}\,\frac{C_g}{f_a}\,,\label{eq:coupling-gd}
\end{eqnarray}
where $m_n$ is the mass of the neutron and $C_{an\gamma} \approx 0.0033$. Note, that for the QCD axion the coupling $g_d$ is proportional to its mass, while it is generally independent of the mass for an axion-like particle.
The quantity $g_d$ appearing in Eq.~\eqref{eq:coupling-gd} could also receive contributions from the coupling of the axion with fermions~\cite{Pospelov:1999ha}. Here, we have focused on the irreducible component that results from the coupling of the axion with the gluon field, so the relation above relies on the other contributions being suppressed. Since laboratory experiments constraining the neutron dipole moment are directly sensitive to $g_d$ in Eq.~\eqref{eq:axion_nucleon_coupling}~\cite{Graham:2013gfa,Baumann:2016wac}, it is useful to also cast our results on the axion-gluon coupling $C_{g}/f_a$ in terms of the axion-EDM coupling $g_d$. It should be however borne in mind that the relation between the two quantities is model dependent, when comparing constraints from cosmological, astrophysical and laboratory probes.

\begin{figure}[ht]
	\centering
	\includegraphics[width=0.7\textwidth]{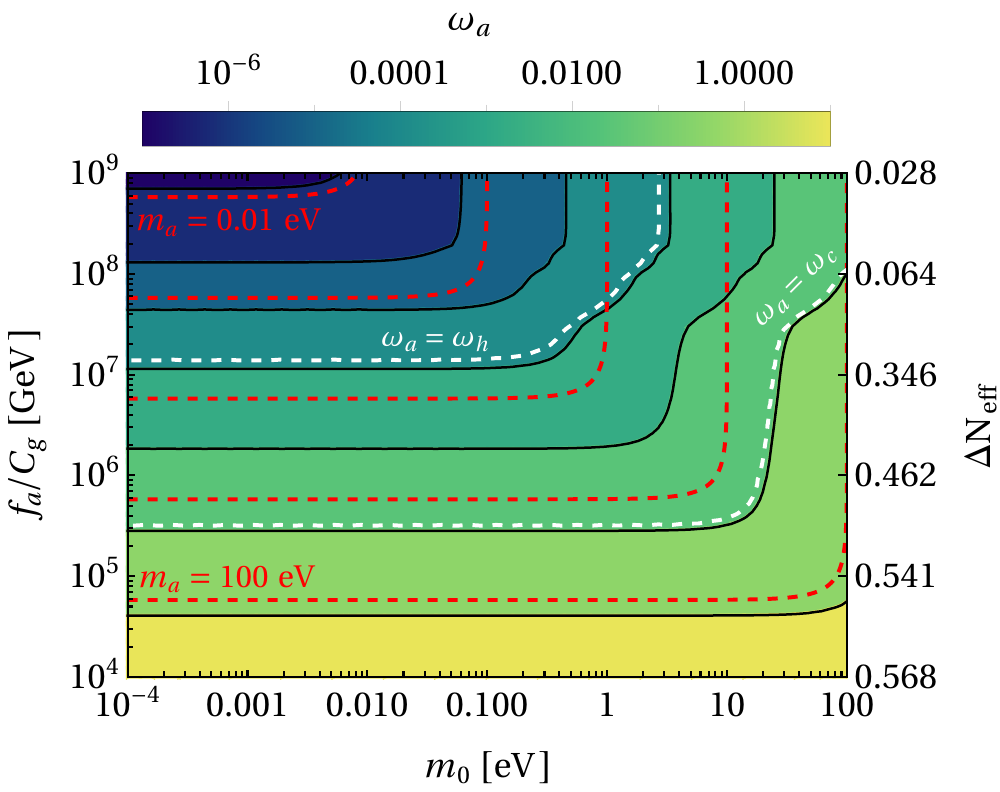}
	\caption{The contour plot shows the axion energy density $\omega_a$ as a function of the explicit breaking axion mass term $m_0$ and the coupling $f_a/C_g$. The dashed red lines show the curves of constant value for the effective axion mass $m_a$ given in Eq.~\eqref{eq:effective-mass}. Also shown are the curves for which either $\omega_a = \omega_h$  or $\omega_a = \omega_c$ (dashed white lines).}
	\label{fig:omega_a}
\end{figure}

The Lagrangian in Eq.~\eqref{eq:lagrangian} can be formally mapped into a chiral axion Lagrangian by a rotation of the quark fields that makes the axion-gluon term disappear~\cite{Georgi:1986df}. Once the effects of the explicit mass breaking are taken into account, the effective axion mass squared in the chiral representation reads~\cite{Georgi:1986df, Aloni:2018vki}
\begin{equation}
    \label{eq:effective-mass}
	m_a^2 = m_0^2 + \left(\frac{C_g}{f_a}\right)^2\,F_\pi^2 m_\pi^2\,\frac{z}{(1+z)^2}\approx m_0^2 + \left(5.8{\rm\,\mu eV}\,\frac{10^{12}{\rm\,GeV}}{f_a/C_g}\right)^2\,,
\end{equation}
where $m_\pi \approx 135\,$MeV and $F_\pi\approx 92\,$MeV are the mass and the decay constant of the neutral pion, respectively, and $z \equiv m_u / m_d \simeq 0.493(19)$~\cite{10.1093/ptep/ptaa104} is the ratio between the masses of the up and down quarks. The effective mass $m_a$ contains two contributions from the explicit symmetry breaking term $m_0$ and from the effective mixing of the axion with the neutral $\pi$, $\eta$, $\eta'$ mesons (see e.g.\ Ref.~\cite{Peccei:2006as}). The second term in the mass squared expression is induced by the axion-gluon coupling and can be lowered with fine-tuning mechanisms such as a tachyonic mass contribution~\cite{Blum:2014vsa}. The presence of $m_0$ prevents the axion from solving the strong-CP problem, and for sufficiently large values of $m_0$ Eq.~\eqref{eq:effective-mass} gives $m_a \approx m_0$ so that the axion mass can be treated as an independent parameter. Here, we consider a regime in which the value of the effective mass transitions from being dominated by the QCD effects to the value $m_0$. Since we restrict the analysis to the region $m_a \gtrsim 0.1\,$meV, QCD effects can be ignored for sufficiently large values of $f_a \gtrsim 10^{10}\,$GeV.
In addition to the effective axion mass, the mapping onto the chiral axion Lagrangian leads to an effective coupling of the axion with the photon, with strength
\begin{equation}
    \label{eq:effective-coupling}
    g_{a\gamma} = g_{a\gamma}^0 - \frac{\alpha_{\rm EM}}{3\pi}\frac{C_g}{f_a}\frac{4+z}{1+z} \approx g_{a\gamma}^0 - 2.3\times 10^{-15}{\rm\,GeV^{-1}}\,\left(\frac{10^{12}{\rm\,GeV}}{f_a/C_g}\right)\,.
\end{equation}
Similarly to the approach for the axion mass $m_a$, in the analysis we focus on the effective coupling $g_{a\gamma}$ which we treat as an independent parameter. Fig.~\ref{fig:omega_a} shows a contour plot of the axion energy density $\omega_a$ over the plane $m_0$ (horizontal axis) and the coupling $f_a/C_g$ (vertical left axis), or equivalently $\Delta N_{\rm eff}$ (vertical right axis), as defined in Eq.~\eqref{eq:axion-density} below. The dashed white lines mark the curves for which either $\omega_a = \omega_h$  or $\omega_a = \omega_c$. Also shown are the curves for which the effective axion mass $m_a$ in Eq.~\eqref{eq:effective-mass} acquires a constant value (dashed red line), as a function of $m_0$ and $f_a/C_g$. The choices ``$m_a = 0.01\,$eV'' and ``$m_a = 100\,$eV'' have been explicitly marked.

\subsection{Relic abundance of thermal axions}

We now turn to describing the present population of thermally-produced axions. The present number density of axions $n_a(T_0)$, where $T_0$ is the temperature of cosmic photons today, is related to the photon number density $n_\gamma$ as
\begin{equation}
    n_a(T_0) = \left(\frac{g_{*s}(T_0)}{g_{*s}(T_d)}\right) \left(\frac{g_a}{g_\gamma}\right) n_\gamma \, ,
\end{equation}
where $g_a = 1$ and $g_\gamma = 2$ are the number of internal degrees of freedom for axions and photons, respectively, and $g_{*s}(T)$ is the effective number of entropy degrees of freedom in the SM as a function of temperature $T$.\footnote{In this work we adopt the parametrization of $g_{*s}(T)$ given in Ref~\cite{Saikawa:2018rcs}, see also~\cite{Drees:2015exa} for a different parametrization and~\cite{DEramo:2021lgb} for a comparison of the two approaches and results. Note in particular that the difference is negligible for decoupling temperatures $T_d \lesssim 100 \, \mathrm{MeV}$ (see Fig.~13 of~\cite{DEramo:2021lgb}) corresponding to the range in reach of current experiments for light axions, see Fig.~\ref{fig:Tdvsma}. Although a higher decoupling temperature can be probed for heavier axions of mass $m_a \gtrsim 1 \, {\rm eV}$, we expect that our choice for the parametrization of $g_{*s}(T)$ would not significantly impact on the final constraints.}
Setting $g_{*s}(T_0) \simeq 3.94$ and $n_\gamma \simeq 411 \, \mathrm{cm}^{-3}$~\cite{Mather:1993ij, Fixsen:1996nj},
the fractional abundance of relativistic axions at present is \MLc{Questa \`e strettamente vera solo per $m_a \gg T_{a,0}$. Per\`o per la nostra analisi non fa differenza (intanto perch\'e la disuguaglianza di prima \`e vera quasi ovunque nel nostro spazio dei parametri; e poi perch\'e, se non ricordo male, nel codice usiamo $\omega_a$ solo per calcolare la massa, quindi \`e come se la nostra fosse una $\omega_a$ ``efficace''). }
\begin{equation}
	\label{eq:axion-density}
	\omega_a \equiv \Omega_a h^2 \simeq \left( \frac{m_a}{130 \, \mathrm{eV}} \right) \left( \frac{g_{*s}(T_d)}{10} \right)^{-1} \, ,
\end{equation}
where $\Omega_a$ is the energy density of axions in units of the critical density $\rho_{\rm crit} = 3H_0^2/(8\pi G_N)$, $H_0$ is the Hubble constant, and $h = H_0/(100{\rm\,km\,s^{-1}\,Mpc^{-1}})$ is the reduced Hubble constant. Relativistic thermal axions in the early Universe contribute to the total energy density of radiation. This is usually parametrized by introducing the effective number of relativistic species $N_\mathrm{eff}$ as
\begin{equation}
	\rho_\mathrm{rad} = \rho_\gamma \left[ 1 + \frac{7}{8}\left(\frac{4}{11}\right)^{4/3} N_\mathrm{eff} \right] \, ,
\end{equation}
where $\rho_\gamma = (\pi^2/15) T^4$ is the energy density of cosmic photons. As mentioned in the introduction,  $N_\mathrm{eff}$ can be decomposed as
\begin{equation}
	N_\mathrm{eff} \equiv N_{\rm eff}^{\rm SM} + \Delta N_\mathrm{eff} \, ,
\end{equation}
where $N_{\rm eff}^{\rm SM} = 3.046$ represents the contribution from the three SM neutrinos~\cite{Mangano:2001iu},\footnote{Recently, the value $N_{\rm eff}^{\rm SM} = 3.044$ has been obtained independently in Refs.~\cite{Bennett:2019ewm,Bennett:2020zkv,Akita:2020szl,Froustey:2020mcq}, where the authors have updated previous calculations taking into account the effects of neutrino flavour oscillations, finite-temperature QED corrections to higher orders in the elementary electric charge, and the complete expression of the collision integrals, while assessing the impact of numerical convergence. Although $N_{\rm eff}^{\rm SM} = 3.044$ is the newly recommended value, the difference is irrelevant given the sensitivity of current experiments, hence we fix the value to $N_{\rm eff}^{\rm SM} = 3.046$ in our analysis.}
\ML{after electron-positron annihilation ($T< 1\,\MeV$}), and $\Delta N_\mathrm{eff}$ quantifies the deviation from the expected value in the standard cosmological model due to, e.g., additional relic species. \ML{For a massless (``mless'') species, and in the absence of entropy production after $e^+e^-$ annihilation, $\DNeff$ does not change with time. In particular, for a massless axion,
$\DNeff$ can be simply obtained as the density of a massless axion relative to the density $\rho_{\nu,\mathrm{mless}}$ of a single  massless neutrino species, i.e.}
\begin{equation}
	\label{eq:Delta-Neff}
	\Delta N_\mathrm{eff} \equiv \frac{\rho_a(m_a=0)}{\rho_{\nu,\mathrm{mless}}} = \frac{\rho_a(m_a=0)}{\frac{7}{8} \left(\frac{4}{11}\right)^{4/3} \rho_\gamma} = \frac{8}{7} \left(\frac{11}{4}\right)^{4/3} \left(\frac{g_a}{g_\gamma}\right) \left(\frac{T_a}{T_\gamma}\right)^4 \, .
\end{equation}
However, the evolution of the axion field takes into account the mass $m_a$ which could not be small compared to the temperature of cosmic photons at the time of recombination $T_{\mathrm{CMB}} \simeq 0.26 \, \mathrm{eV}$. \ML{Therefore, as explained at the beginning of the section, we find it useful to consider $\Delta N_\mathrm{eff}^{\rm CMB} \equiv \rho_a/\rho_{\nu,\mathrm{mless}}$. This is related to $\Delta N_\mathrm{eff}$ through}
\begin{equation}
    \label{eq:DNeff-CMB}
    \Delta N_\mathrm{eff}^{\rm CMB} = \frac{15}{\pi^4} \Delta N_\mathrm{eff} \int_0^\infty \frac{y^2\sqrt{y^2+m^2_a/T^2_\mathrm{CMB}}}{e^y-1}dy \,.
\end{equation}
Note that the axion mass does not enter the distribution function in Eq.~\eqref{eq:Delta-Neff} and Eq.~\eqref{eq:DNeff-CMB} since axions decouple from the primordial plasma while they are still relativistic.

For small values of $m_a$, the two quantities in Eq.~\eqref{eq:Delta-Neff} and Eq.~\eqref{eq:DNeff-CMB} lead to the same results, as shown from the behaviour of the solid cyan line in Fig.~\ref{fig:Tdvsma}, which approaches a horizontal line as the axion mass decreases. In the limit of a massless axion, the result $\Delta N_\mathrm{eff} < 0.376$ at 95\% CL corresponds to the bound obtained for the lightest axion mass considered in the run, $m_a = 10^{-4} \, \mathrm{eV}$, see also Fig.~\ref{fig:Tdvsma}. For heavier masses, the bound relaxes and disappears completely for $m_a \gtrsim 0.1\,$eV where the axion ceases to behave as a massless particle. At the same time, the cyan line represents the constraint obtained directly from the analysis of the different runs with a fixed axion mass. For axion masses $m_a \lesssim 0.1 \, \mathrm{eV}$ the constraints are mainly due to the direct bound on $\Delta N_\mathrm{eff}^{\rm CMB}$ at recombination and join those obtained for a massless axion, whereas for axion masses $m_a \gtrsim 0.1 \, \mathrm{eV}$ the bounds are mainly due to the constraints on the axion relic abundance $\omega_a$ (see also the discussion in Section~\ref{sec:results_DNeff}).

After axions decouple from the primordial plasma, particles becoming nonrelativistic release their entropy in the CMB photons, while the axion temperature is unaffected. Therefore, using entropy conservation we can write the ratio of the axion and photon temperatures as
\begin{equation}
	\label{eq:Ta}
	\frac{T_a}{T_\gamma} = \left( \frac{g_{*s}(T_{\mathrm{CMB}})}{g_{*s}(T_d)} \right)^{1/3} \, ,
\end{equation}
where $g_{*s}(T_{\mathrm{CMB}}) = 2 + (7/11)N_{\rm eff}^{\rm SM} \simeq 3.94$. Inserting Eq.~\eqref{eq:Ta} into Eq.~\eqref{eq:Delta-Neff} gives
\begin{equation}
	\label{eq:DNeff}
	\Delta N_\mathrm{eff} 
	\simeq 0.027 \left( \frac{g_{*s}(T_d)}{106.75} \right)^{-4/3} \, ,
\end{equation}
so that the fractional axion abundance as a function of $\Delta N_\mathrm{eff}$ results in
\begin{equation}
	\label{eq:omega_a_DNeff}
	\omega_a \simeq 0.011 \left( \frac{m_a}{\mathrm{eV}} \right) \Delta N_\mathrm{eff}^{3/4} \, .
\end{equation}

To have a better understanding of the CMB constraints on axions, it is important to determine whether axions behave as a hot or cold DM component at the epoch of CMB decoupling. An estimate is obtained by computing the ratio between the thermally-averaged momentum and the axion mass at recombination, which is given by~\cite{Carenza:2021ebx}
\begin{equation}
    \label{eq:effectivemomentum}
    \frac{\langle p_{a,\mathrm{rec}} \rangle}{m_a} \approx 2.7 \, \frac{T_{a,\mathrm{rec}}}{m_a} = 2.7 \, \frac{T_\mathrm{CMB}}{m_a} \left(\frac{g_{*s}(T_\mathrm{CMB})}{g_{*s}(T_d)}\right)^{1/3} \, .
\end{equation}
At the epoch of recombination, axions are hot DM provided $\langle p_{a,\mathrm{rec}} \rangle / m_a \gtrsim 1$. In the opposite limit $\langle p_{a,\mathrm{rec}} \rangle / m_a \ll 1$, axions behave as a CDM component. In the first case, axions are relativistic at recombination and the parameter space is bound directly by the CMB constraints on $N_{\rm eff}$. In the second case where axions do not contribute to $N_\mathrm{eff}$ and are nonrelativistic at recombination, the parameter space is constrained by requiring that the relic abundance of thermal axions does not exceed that of CDM, namely $\omega_a \le \omega_c$.

\section{Axion decoupling from the primordial plasma}
\label{sec:decoupling}

The ultimate goal of this work is to constrain the axion mass and its couplings to the photon and gluon fields. To this extent, it is necessary to have a relation between these couplings and the decoupling temperature $T_d$, which is in turn related to $\Delta N_\mathrm{eff}$ through Eq.~\eqref{eq:DNeff}. This is provided by introducing the thermal axion production rate $\gamma \equiv {\rm d}n_a/{\rm d}t$ which expresses the rate of production of thermal axions from the primordial plasma times the number density of axions $n_a$, as derived from the imaginary part of the axion self-energy~\cite{Weldon:1990iw, Gale:1990pn, Raffelt:1996wa, Masso:2002np}. The evolution of the axion number density $n_a = n_a(T)$ proceeds according to a Boltzmann equation,
\begin{equation}
    \frac{{\rm d}n_a}{{\rm d}t} + 3H(T)n_a = \gamma\left(1 - \frac{n_a}{n_a^{\rm eq}}\right)\,,
\end{equation}
where $n_a^{\rm eq}(T) = (\zeta(3)/\pi^2)g_aT^3$ is the number density at equilibrium. Freeze-out occurs when the axion production rate ceases to keep pace with the expansion rate, at the decoupling temperature
\begin{equation}
    \label{eq:decoupling}
    H(T_d) \simeq \Gamma \equiv \gamma/n_a^{\rm eq}(T_d)\,.
\end{equation}
The relation between $T_d$ and $f_a$ is provided by the expression above. This is the subject of the following subsections, where the two cases for the axion-gluon and axion-photon couplings are discussed separately.

\subsection{Axion-gluon coupling}

The axion-gluon coupling given by the Lagrangian term $\mathcal{L}_{ag}$ is responsible for the production of axions in the early Universe from the quark-gluon plasma. 
For the coupling of the axion to gluons, the production rate is~\cite{Graf:2010tv}
\begin{equation}
    \label{eq:gamma_g}
    \gamma_g = \frac{\zeta(3)}{4\pi^5}\,\alpha_s^2\,T^6\,\left(\frac{C_g}{f_a}\right)^2\,F_g(T)\,,
\end{equation}
where $\zeta(3) \approx 1.202$ and the function $F_g(T)$ encodes the production of axions through gluons and other SM particles~\cite{Salvio:2013iaa}. The suffix ``$g$'' in the rate refers to the production specifically from the quark-gluon plasma through the axion-gluon coupling.

The axion production rate is generally computed in regimes far away from the QCD phase transition, at which non-trivial dynamics occurs. At a temperature $T_N \simeq 2 \, \mathrm{GeV}$ well above the QCD phase transition,\footnote{At this temperature the strong coupling constant is $\alpha_s(T_N) \simeq 0.3$, thus for $T<T_N$ a perturbative treatment of QCD processes is no longer justified.} the axion production rate is regulated by gluon scattering processes ($g + g \rightarrow g + a$), quark/antiquark annihilations ($q + \bar{q} \rightarrow g + a$) and scatterings between quarks/antiquarks and gluons ($q/\bar{q} + g \rightarrow q/\bar{q} + a$). Instead, at temperatures below the QCD phase transition, the leading contribution is given by pion scatterings, namely $\pi^+ \pi^- \rightarrow \pi^0 + a$ and $\pi^+/\pi^- + \pi^0 \rightarrow \pi^+/\pi^- + a$. As shown in~\cite{DiLuzio:2021vjd}, the calculation for the axion-pion scattering rates using chiral perturbation theory (ChPT) is not reliable above the temperature $T_\mathrm{ChPT} \simeq 62 \, \mathrm{MeV}$, since at higher temperatures perturbations can no longer be neglected and the effective field theory description breaks down. To overcome this issue, the axion production rate between the temperatures $T_\mathrm{ChPT}$ and $T_N$ has been obtained in Refs.~\cite{DEramo:2021psx,DEramo:2021lgb} by interpolating between the two regimes. The results in Refs.~\cite{DEramo:2021psx,DEramo:2021lgb} bridge between these two regimes and provide us with the most up-to-date description for the function $\gamma_g f_a^2$, as reported in Figure 2 of Ref.~\cite{DEramo:2021psx}, where the quantity $\gamma_g f_a^2$ is shown as a function of the temperature of the primordial plasma.

\begin{figure}
	\centering
	\includegraphics[width=0.45\textwidth]{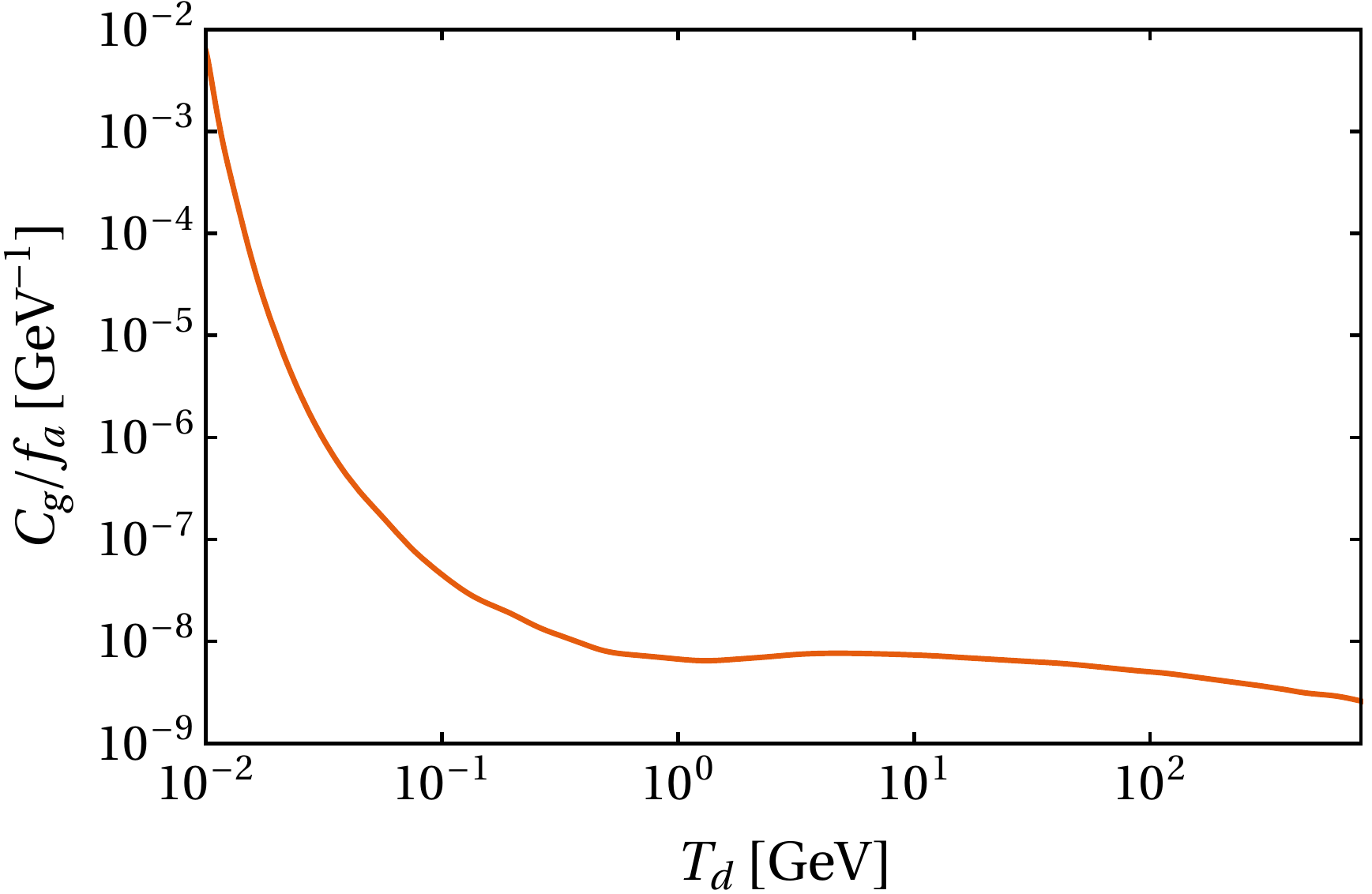}
	\includegraphics[width=0.45\textwidth]{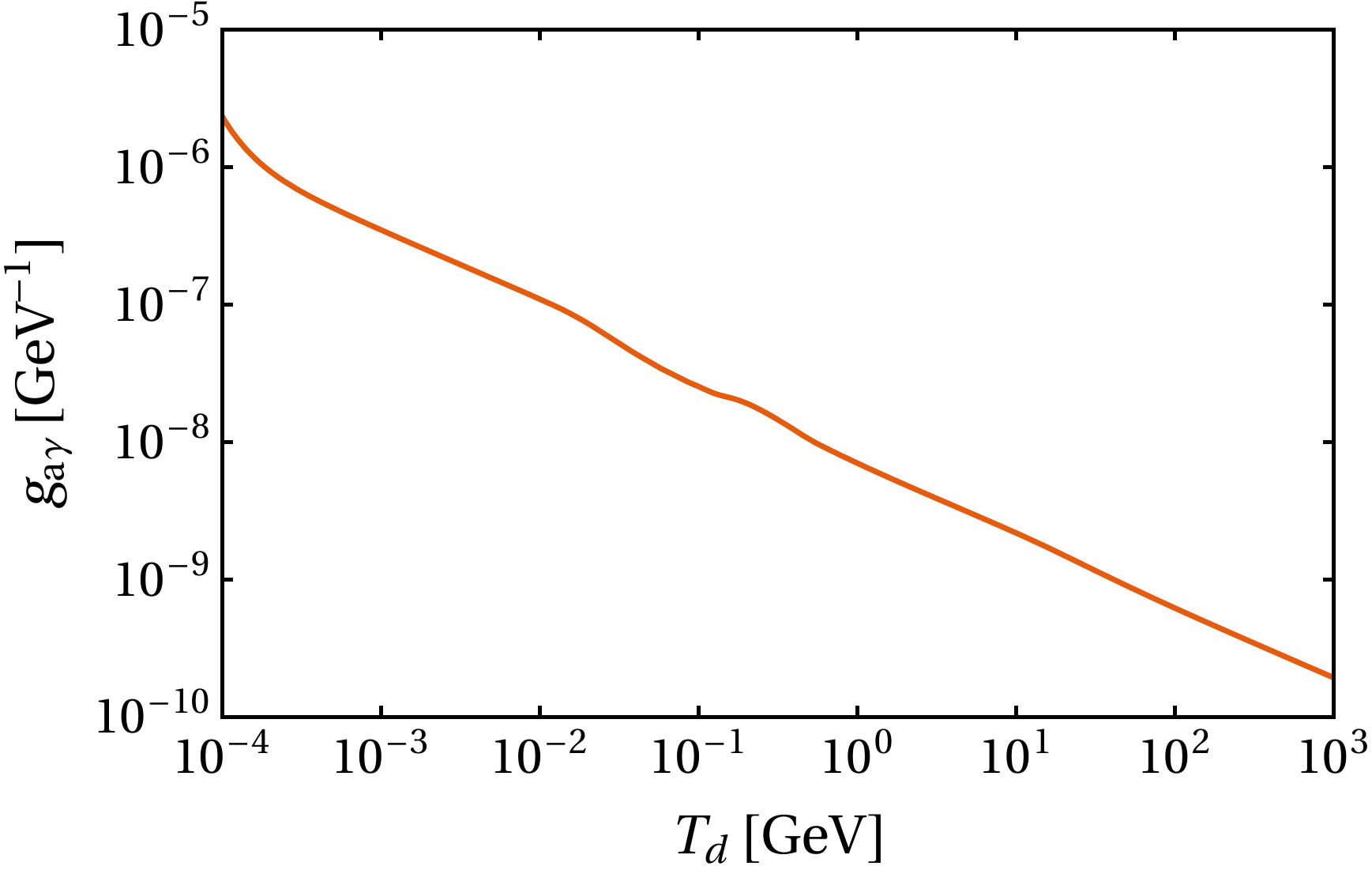}
	\caption{\textit{Left panel}: Axion-gluon coupling, parametrized by the inverse axion decay constant $C_g / f_a$, as a function of the decoupling temperature $T_d$. The freeze-out condition in Eq.~\eqref{eq:decoupling} is imposed together with the results in Figure 2 of Ref.~\cite{DEramo:2021psx} to obtain $T_d$. \textit{Right panel}: Axion-photon coupling $g_{a\gamma}$ as a function of the decoupling temperature $T_d$, as described in Eq.~\eqref{eq:axion-photon-coupling}.}
	\label{fig:fa_Td}
\end{figure}

Here, we adopt the numerical results obtained in Ref.~\cite{DEramo:2021psx} for the function $\gamma_g f_a^2$ and, assuming the standard cosmological model holds for the Hubble rate $H(T)$ at temperature $T$, we impose the condition of thermal production at freeze-out in Eq.~\eqref{eq:decoupling} to obtain a relation between the axion decay constant $f_a/C_g$ and the decoupling temperature $T_d$. Figure~\ref{fig:fa_Td} (left panel) shows the result obtained for this relation over the range of temperatures considered in this work. This is the relation we use to interpret the results of the Monte Carlo in Sec.~\ref{sec:results} and relate them with the axion production from the quark-gluon plasma to constrain the coupling $C_g/f_a$, or the EDM coupling $g_d$ in Eq.~\eqref{eq:coupling-gd}.

\subsection{Axion-photon coupling}

\begin{figure}
	\centering
	\includegraphics[width=0.7\textwidth]{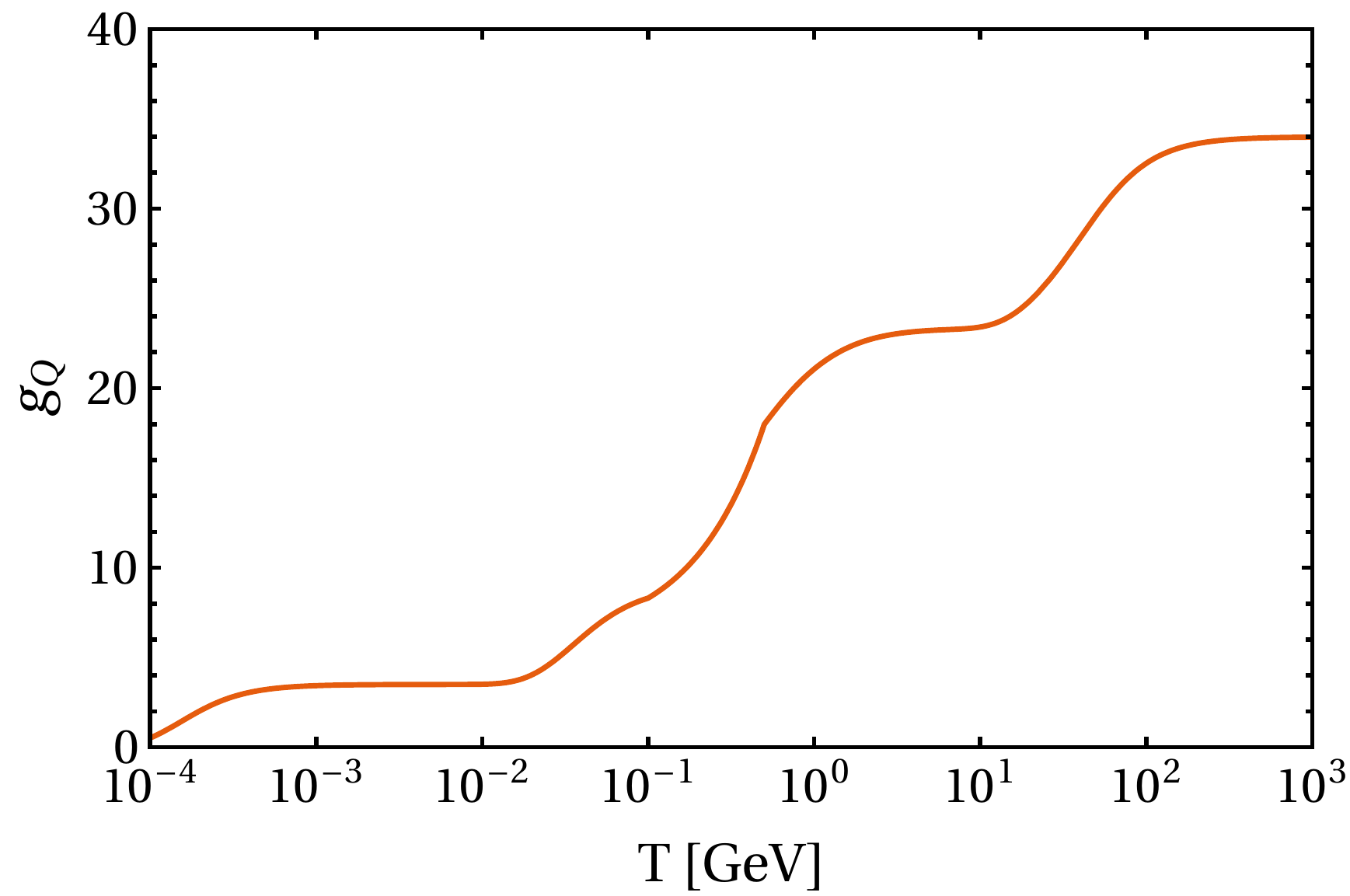}
	\caption{The effective number of charged relativistic degrees of freedom $g_Q(T)$ as a function of the temperature of the primordial plasma $T$.}
	\label{fig:g_starQ}
\end{figure}

The leading mechanism contributing to the production of thermal axions due to an axion-photon coupling is the Primakoff effect, which describes the resonant conversion of photons into axions in the presence of the strong magnetic field of charged particles. Production in the primordial plasma proceeds with the rate~\cite{Bolz:2000fu,Cadamuro:2011fd,Millea:2020xxp}
\begin{equation}
    \label{eq:primakoff}
	\Gamma_{Q\gamma \rightarrow Qa} \simeq \frac{\alpha_{\rm EM} \pi^2 g^2_{a\gamma}}{36\zeta(3)} \left[ \ln\left(\frac{T^2}{m_\gamma^2}\right) + 0.8194 \right] n_Q \, ,
\end{equation}
where $m_\gamma = T/(6\alpha_{\rm EM} \sqrt{g_Q(T)})$ is the plasmon mass, $g_Q(T) = \sum_i Q_i^2 g_{*,i}(T)$ is the effective number of relativistic degrees of freedom for the $i$th charged species of charge $Q_i$, and $n_Q = \sum_i Q_i^2 n_i \equiv (\zeta(3)/\pi^2) g_Q(T) T^3$ is the effective number density of charged particles in the cosmological plasma.
In principle, to derive the evolution of $g_Q(T)$ with the temperature we would need to isolate the contribution to $g_*(T)$ from each charged particle species and weight it by the particle's charge squared.
This is particularly challenging during the QCD phase transition, where the total contribution to $g_*(T)$ is usually derived from lattice QCD results~\cite{Drees:2015exa,HotQCD:2014kol,Saikawa:2018rcs}. To overcome this issue, we first compute $g_Q$ for temperatures $T < 100 \, \mathrm{MeV}$ and $T > 500 \, \mathrm{MeV}$. 
The behavior of $g_Q$ within the temperature range $100 \, \mathrm{MeV} < T < 500 \, \mathrm{MeV}$, where the QCD phase transition takes place, is then reconstructed by interpolating between the two regimes previously obtained. This procedure is similar to the method used in Ref.~\cite{DEramo:2021psx} to reconstruct the evolution of $\gamma_gf_a^2$. The result is shown in Fig.~\ref{fig:g_starQ} for the whole range of temperatures considered. The Primakoff effect establishes a thermal population of axions that decouples from the primordial plasma at the temperature $T_d$, which is related to the axion-photon coupling $g_{a\gamma}$ by~\cite{Cadamuro:2011fd}
\begin{equation}
    \label{eq:axion-photon-coupling}
    g_{a\gamma} \simeq 10^{-8} \times \frac{\sqrt{g_*}}{g_Q} \left( \frac{T_d}{\mathrm{GeV}} \right)^{-1} \, \mathrm{GeV}^{-1} \, .
\end{equation}

In addition to the thermal production just discussed, a non-zero coupling $g_{a\gamma}$ would also induce the axion to decay into a pair of photons, with a rate given by (see e.g.\ Refs.~\cite{Cadamuro:2010cz,Cadamuro:2011fd,Conlon:2013isa})
\begin{equation}
    \label{eq:decayrate}
    \Gamma_{a \rightarrow \gamma\gamma} = \frac{g_{a\gamma}^2 m_a^3}{64\pi} \, .
\end{equation}
The axion is a stable relic over cosmological timescales if the decay time is larger than the age of the Universe, $\tau_D \sim \Gamma_{a \rightarrow \gamma\gamma}^{-1} \gg H_0^{-1}$. Inserting the expression in Eq.~\eqref{eq:decayrate} leads to the requirement
\begin{equation}
	\label{eq:axion-decay}
	\frac{\Gamma_{a \rightarrow \gamma\gamma}}{H_0} \simeq 3.48 \times 10^{-2} \left(\frac{g_{a\gamma}}{10^{-7} \, \mathrm{GeV}^{-1}}\right)^2 \left(\frac{m_a}{\mathrm{eV}}\right)^3 \ll 1 \, .
\end{equation}
The decay of axions would be accompanied by a reduction of their cosmological abundance and an injection of photons with energy $E_\gamma = m_a/2$. Since a dedicated analysis is required in this scenario, we do not consider this region of the parameter space in the results and we impose the bound in Eq.~\eqref{eq:axion-decay} in the analysis.

\section{Data sets and analysis}
\label{sec:analysis}

We now discuss the data sets composition and the software used in the analysis. We use the most recent {\it Planck} 2018 data of CMB temperature and polarization anisotropies~\cite{Planck:2018vyg} together with BAO data from galaxy surveys, consisting of BOSS DR12~\cite{BOSS:2016wmc}, 6dFGS~\cite{Beutler:2011hx} and SDSS-MGS~\cite{Ross:2014qpa}. To compute the theoretical predictions we employ a modified version of the publicly-available Boltzmann solver code {\ttfamily{CAMB}}, which correctly includes the propagation of the axion by incorporating a Bose-Einstein distribution function.\footnote{This code has been previously used in Ref.~\cite{Carenza:2021ebx} to derive the cosmological constraints on thermal QCD axions in low-reheating scenarios.} This is an advance compared with the majority of previous approaches, in which cosmological constraints on bosonic thermal relics had been obtained by treating them as additional effective neutrino species (see instead Refs.~\cite{Archidiacono:2019wdp,Carenza:2021ebx} for a few exceptions). The Markov Chain Monte Carlo sampler {\ttfamily{CosmoMC}}~\cite{Lewis:2002ah} is then used to derive constraints on the parameters of the model. The {\ttfamily{CosmoMC}} code consistently evaluates the helium mass fraction $Y_p$ as a function of $N_{\rm eff}$ and the baryon density parameter using the interpolation grids from Parthenope~\cite{Pisanti:2007hk} and PRIMAT~\cite{Pitrou:2018cgg}, assuming standard Big Bang Nucleosynthesis (BBN). This value of $Y_p$ is then fed to {\ttfamily{CAMB}} when computing the theoretical power spectrum of CMB anisotropies. We check the convergence of the chains by controlling that the Gelman-Rubin parameter $R-1 < 0.01$, with the first 30\% of total steps discarded as burn-in. The posterior distributions are then obtained and plotted using the {\ttfamily{GetDist}} package~\cite{Lewis:2019xzd}.

Following the discussions in the previous sections, we extend the standard cosmological model (or $\Lambda$CDM model) with two additional parameters:\footnote{We stress again that for a specific QCD axion model, the mass and decoupling temperature are related via Eq.~\eqref{eq:effective-mass} with $m_0 = 0$ and $C_g = 1$. Hence, in the case of the QCD axion we would have only one additional parameter with respect to $\Lambda$CDM, instead of the two parameters used here, since $g_{a\gamma}\propto m_a\propto 1/f_a$.} the axion mass $m_a$ and its contribution to the effective number of relativistic species, $\Delta N_\mathrm{eff}$. We refer to this as the $\Lambda$CDM+$\Delta N_\mathrm{eff}$+$m_a$ model. The vector spanning the eight-dimensional parameter space is represented by $\Theta_{\Delta N_\mathrm{eff}+m_a} = \{ \omega_b,\, \theta_s, \, \tau, \, \ln(10^{10}A_s), \, n_s,\, \omega_{c+a},\, m_a, \, \Delta N_\mathrm{eff} \}$, where $\omega_b \equiv \Omega_b h^2$ is the physical density of baryons, $\theta_s$ is the angular acoustic scale at recombination, $\tau$ is the reionization optical depth, $A_s$ is the normalization of the power spectrum, and $n_s$ is the scalar spectral index, see Ref.~\cite{Planck:2018vyg} for additional information. In addition, $\omega_{c+a} \equiv \omega_c + \omega_a$ labels the density of CDM plus axions. The latter parametrization is useful because for certain values of $m_a$ thermal axions behave as CDM, so that in this case $\omega_{c+a}$ quantifies the total amount of cold particles. The ``non-axionic'' CDM abundance can then be derived as $\omega_c = \omega_{c+a} - \omega_a$, where $\omega_a$ is computed via Eq.~\eqref{eq:omega_a_DNeff}. Given the modification on $\omega_{c+a}$, the first six parameters in the vector $\Theta_{\Delta N_\mathrm{eff}+m_a}$ describe the $\Lambda$CDM model. The priors for each component of $\Theta_{\Delta N_\mathrm{eff}+m_a}$ are reported in Table~\ref{tab:priors}. In particular, we draw the values of $\Delta N_\mathrm{eff}$ from a flat prior $\Delta N_\mathrm{eff} \in [0,1]$, for the axion mass we set a logarithmic prior as $\log_{10} (m_a/\mathrm{eV}) \in [-6,4]$, while the list of flat priors on the other cosmological parameters are chosen so that the results of $\Lambda$CDM are recovered when the axion field is removed. We assume that only one of the active neutrino species is massive, and we fix the sum of neutrino masses to the minimal value allowed by flavour oscillation experiments in the normal hierarchy scenario, $\sum m_\nu = 0.06 \, \mathrm{eV}$. Once the \texttt{CAMB} model is run with this setup and the model has been constrained, the bounds on $\Delta N_\mathrm{eff}$ are then converted into constraints on the axion couplings to photons and gluons using using Eq.~\eqref{eq:DNeff} to relate $\Delta N_\mathrm{eff}$ with the decoupling temperature $T_d$. Results for the axion couplings are shown in Sec.~\ref{sec:results_ma_DNeff} below, see Eq.~\eqref{eq:axion-photon-coupling} and the relation plotted in Fig.~\ref{fig:fa_Td}.
\begin{table}
	\centering
	\begin{tabular}{ |c|c| } 
		\hline
		\textbf{Parameter} & \textbf{Prior} \\
		\hline
		\hline
		$\omega_b$ & $[0.005,0.1]$ \\[2mm]
		$\theta_s$ & $[0.5,10]$ \\[2mm]
		$\tau$ & $[0.01,0.8]$ \\[2mm]
		$\ln(10^{10} A_s)$ & $[1.61,3.91]$ \\[2mm]
		$n_s$ & $[0.8,1.2]$ \\[2mm]
		$\omega_{c+a}$ & $[0.001,0.99]$ \\[2mm]
		$\Delta N_\mathrm{eff}$ & $[0,1]$ \\[2mm]
		$\log_{10}(m_a/\mathrm{eV})$ & $[-6,4]$ \\
		\hline
	\end{tabular}
	\caption{Priors used for the cosmological parameters in the $\Lambda$CDM+$\Delta N_\mathrm{eff}$+$m_a$ model. For the $\Lambda$CDM+$\Delta N_\mathrm{eff}$ the same priors are used, except for the axion mass which is fixed to a specific value.}
	\label{tab:priors}
\end{table}

We also perform a series of runs, here $\Lambda$CDM+$\Delta N_\mathrm{eff}$, in which the mass of the axion is fixed and we allow $\Delta N_\mathrm{eff}$ to vary over the range in Table~\ref{tab:priors}, thus implementing the runs over the parameter space spanned by the vector $\Theta_{\Delta N_\mathrm{eff}} \equiv \{ \omega_b,\, \theta_s, \, \tau, \, \ln(10^{10}A_s), \, n_s,\, \omega_{c+a},\, \Delta N_\mathrm{eff} \}$. For each of these runs, we fix the axion mass to the value chosen within
$$m_a \in \{10^{-4},\, 10^{-3},\, 10^{-2},\, 0.1,\, 1,\, 3,\, 10,\, 30,\, 100\} \, \mathrm{eV}\,.$$
This procedure allows us to determine an upper bound on $\Delta N_\mathrm{eff}$, and ultimately on the axion couplings, as a function of the axion mass. These results have been used to picture the constraints on $\Delta N_{\rm eff}$ in Fig.~\ref{fig:Tdvsma} and they are reported in Sec.~\ref{sec:results_DNeff}.

\section{Results and discussion}
\label{sec:results}

In the following sections we report and discuss the main results of our analysis, considering both cases for the QCD axion and a generic axion-like particle.

\subsection{\texorpdfstring{$\Lambda$CDM+$\Delta N_\mathrm{eff}$+$m_a$}{H0}}
\label{sec:results_ma_DNeff}

\begin{figure}
	\centering
	\includegraphics[width=0.7\textwidth]{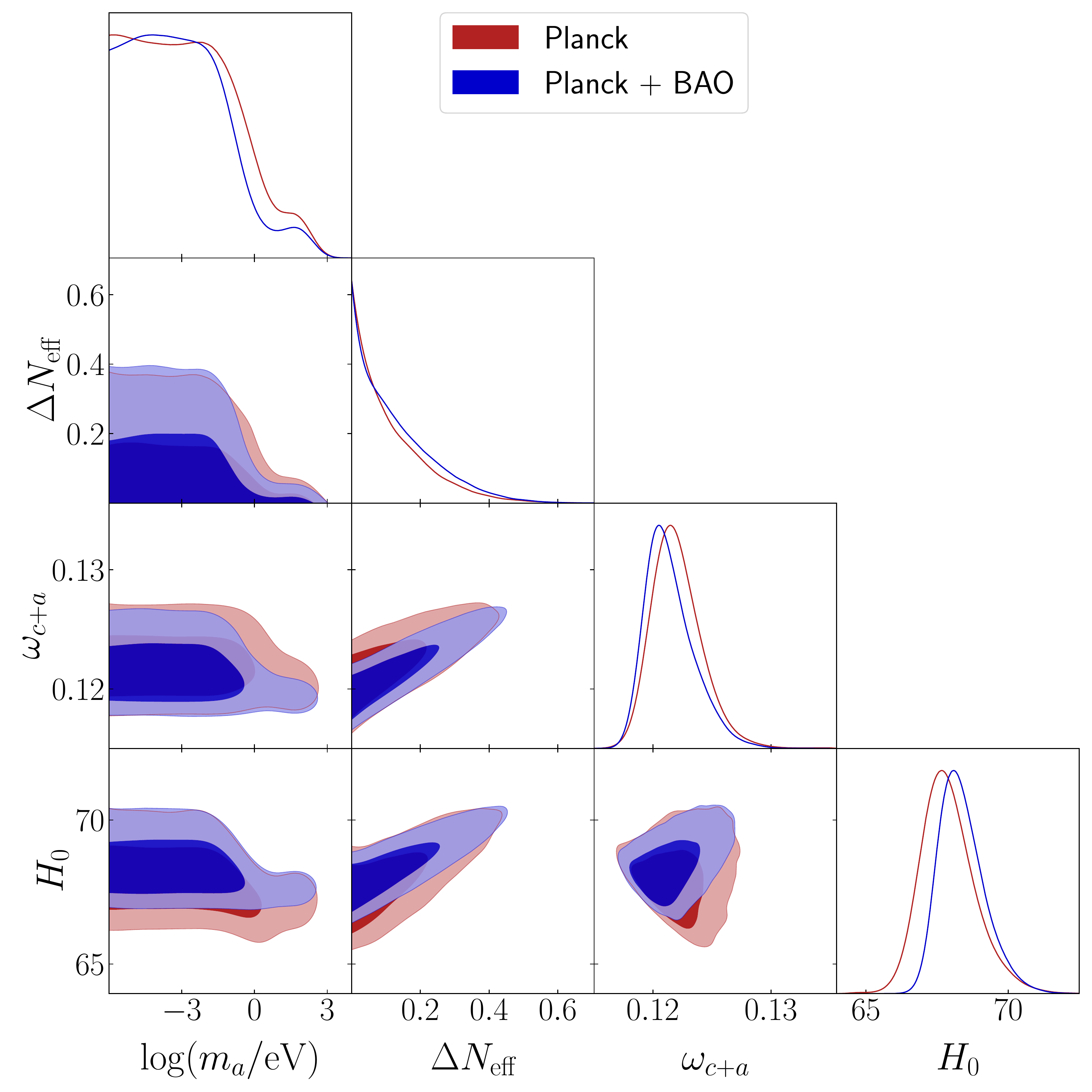}
	\caption{Triangle plot including 2D and 1D posteriors for different cosmological parameters in the $\Lambda$CDM+$\Delta N_\mathrm{eff}$+$m_a$ model. Note in particular the shape of the 1D distribution of $\log(m_a)$ and the impact of including BAO data in the analysis. See text for details.}
	\label{fig:Triangle-DNeff-ma}
\end{figure}

We first discuss the results for the run with the full set of parameters spanned by $\Theta_{\Delta N_{\rm eff}+m_a}$. The corresponding constraints on the cosmological parameters of the $\Lambda$CDM+$\Delta N_\mathrm{eff}$+$m_a$ model are reported in Table~\ref{tab:constraints}. Figure~\ref{fig:Triangle-DNeff-ma} shows the constraints obtained using {\it Planck} 2018 TT,TE,EE+lowE data (red), and with the additional inclusion of BAO data (blue). Note in particular the second peak of the probability distribution of $m_a$ and the degeneracy of $N_\mathrm{eff}$ with $H_0$ and $\omega_{c+a}$. The shape of the 1D distribution of $m_a$ can be understood as follows. At smaller masses, the axion is fully relativistic at the CMB epoch and the constraints mostly come from the axion contribution to $N_\mathrm{eff}$. For very small masses, say $m_a<10^{-2}\,\mathrm{eV}$, the contribution to $\Delta N_\mathrm{eff}$ is close to $\Delta N_\mathrm{eff}^{\rm CMB}$, so that the distribution for low masses is flat. For intermediate masses $10^{-2}\,\mathrm{eV}\lesssim m_a \lesssim 1\,\mathrm{eV}$, the axion must decouple early enough so that its contribution to the energy density budget is sufficiently diluted. This explains the sharp cutoff in the 1D distribution. At higher masses, the axion effectively behaves as a cold dark matter component. The second peak in the 1D distribution corresponds to values of the axion masses that satisfy the cosmological constraints on the abundance of cold dark matter.

The inclusion of BAO data better constrains the late-time dynamics of the Universe, thus sharpening the second peak in the 1D distribution of $m_a$. In more detail, BAO data exclude the high-$\omega_{c+a}$/low-$H_0$ region of the parameter space (see the corresponding 2D contours in Fig.~\ref{fig:Triangle-DNeff-ma}). This effect, combined with the degeneracy between the Hubble constant and $\Delta N_\mathrm{eff}$, explains the behaviour of the 1D distribution of $\Delta N_\mathrm{eff}$ when BAO data are included in the analysis (for a detailed discussion, see next subsections). As obtained in previous literature, we note that hot axions may feebly mitigate the tension over the different Hubble constant measurements (see Ref.~\cite{DiValentino:2021izs} for a recent review) although they do not erase it~\cite{DEramo:2018vss}.

As discussed in Sec.~\ref{sec:decoupling}, the bounds on $\Delta N_{\rm eff}$ can be converted into bounds on $g_{a\gamma}$ using Eq.~\eqref{eq:axion-photon-coupling} when the axion is mainly produced by axion-photon scatterings, or equivalently on a bound on $C_g/f_a$ or $g_d$ using Eq.~\eqref{eq:decoupling} when the axion is mainly produced from quark-gluon interactions. In the range of axion masses spanned in this work, we obtain the following marginalised constraints at 95\% CL on the axion couplings:
\begin{equation}
    g_{a\gamma} <
    \begin{cases}
    2.84 \times 10^{-8}\,\mathrm{GeV^{-1}}\,\quad &\mathit{Planck\;} 2018\\
    3.16 \times 10^{-8}\,\mathrm{GeV^{-1}}\,\quad &\mathit{Planck\;} 2018 + \mathrm{BAO}
    \end{cases}
\end{equation}
\begin{equation}
    \frac{C_g}{f_a} <
    \begin{cases}
    6.98 \times 10^{-8}\,\mathrm{GeV^{-1}}\\
    1.06 \times 10^{-7}\,\mathrm{GeV^{-1}} 
    \end{cases} 
     {\rm or} \quad
    g_d <
    \begin{cases}
    2.47 \times 10^{-10}\,\mathrm{GeV^{-2}}\,\quad &\mathit{Planck\;} 2018\\
    3.77 \times 10^{-10}\,\mathrm{GeV^{-2}}\,\quad &\mathit{Planck\;} 2018 + \mathrm{BAO}
    \end{cases}
\end{equation}

\begin{table}
\def\arraystretch{1.5}
	\centering
	\begin{tabular}{ |c |c |c| } 
		\hline
		Parameter & {\it Planck} TT,TE,EE+lowE & {\it Planck} TT,TE,EE+lowE+BAO \\
		\hline
		\hline
		$\omega_b$ & $0.02242\pm 0.00017$ & $0.02249\pm 0.00015$ \\[2mm]
		$H_0$ & $67.90^{+0.75}_{-1.1}{\rm\,km\,s^{-1}\,Mpc^{-1}}$ & $68.38^{+0.58}_{-0.94}{\rm\,km\,s^{-1}\,Mpc^{-1}}$ \\[2mm]
		$\tau$ & $0.0553^{+0.0074}_{-0.0082}$ & $0.0563\pm 0.0079$ \\[2mm]
		$\ln(10^{10} A_s)$ & $3.051^{+0.016}_{-0.017}$ & $3.051\pm 0.017$ \\[2mm]
		$n_s$ & $0.9676^{+0.0050}_{-0.0061}$ & $0.9700^{+0.0044}_{-0.0053}$ \\[2mm]
		$\omega_{c+a}$ & $0.1221^{+0.0016}_{-0.0024}$ & $0.1215^{+0.0014}_{-0.0025}$ \\[2mm]
		$\Delta N_\mathrm{eff}$ & $< 0.337$ & $<0.358$ \\[2mm]
		$m_a$ & $< 11.66\,$eV & $< 3.14\,$eV \\[2mm]
		\hline
		$g_{a\gamma}$ & $< 2.84 \times 10^{-8} \; \mathrm{GeV}^{-1}$ & $< 3.16 \times 10^{-8} \; \mathrm{GeV}^{-1}$\\[2mm]
		$C_g/f_a$ & $< 6.98 \times 10^{-8} \; \mathrm{GeV}^{-1}$ & $< 1.06 \times 10^{-7} \; \mathrm{GeV}^{-1}$\\[2mm]
		$g_d$ & $< 2.47 \times 10^{-10} \; \mathrm{GeV}^{-2}$ & $< 3.77 \times 10^{-10} \; \mathrm{GeV}^{-2}$\\ 
		\hline
	\end{tabular}
	\caption{Constraints on the cosmological parameters for the $\Lambda$CDM+$\Delta N_\mathrm{eff}$+$m_a$ model, obtained both from using {\it Planck} 2018 TT,TE,EE+lowE datasets alone and in combination with 
	BAO. The limits on the first six parameters are reported at 68\% CL, whereas the upper bounds on $m_a$, $\Delta N_\mathrm{eff}$, and the couplings are at 95\% CL.}
	\label{tab:constraints}
\end{table}

\subsection{\texorpdfstring{$\Lambda$CDM+$\Delta N_\mathrm{eff}$}{H0} with fixed axion mass}
\label{sec:results_DNeff}

\begin{figure}
	\centering
	\includegraphics[width=0.45\textwidth]{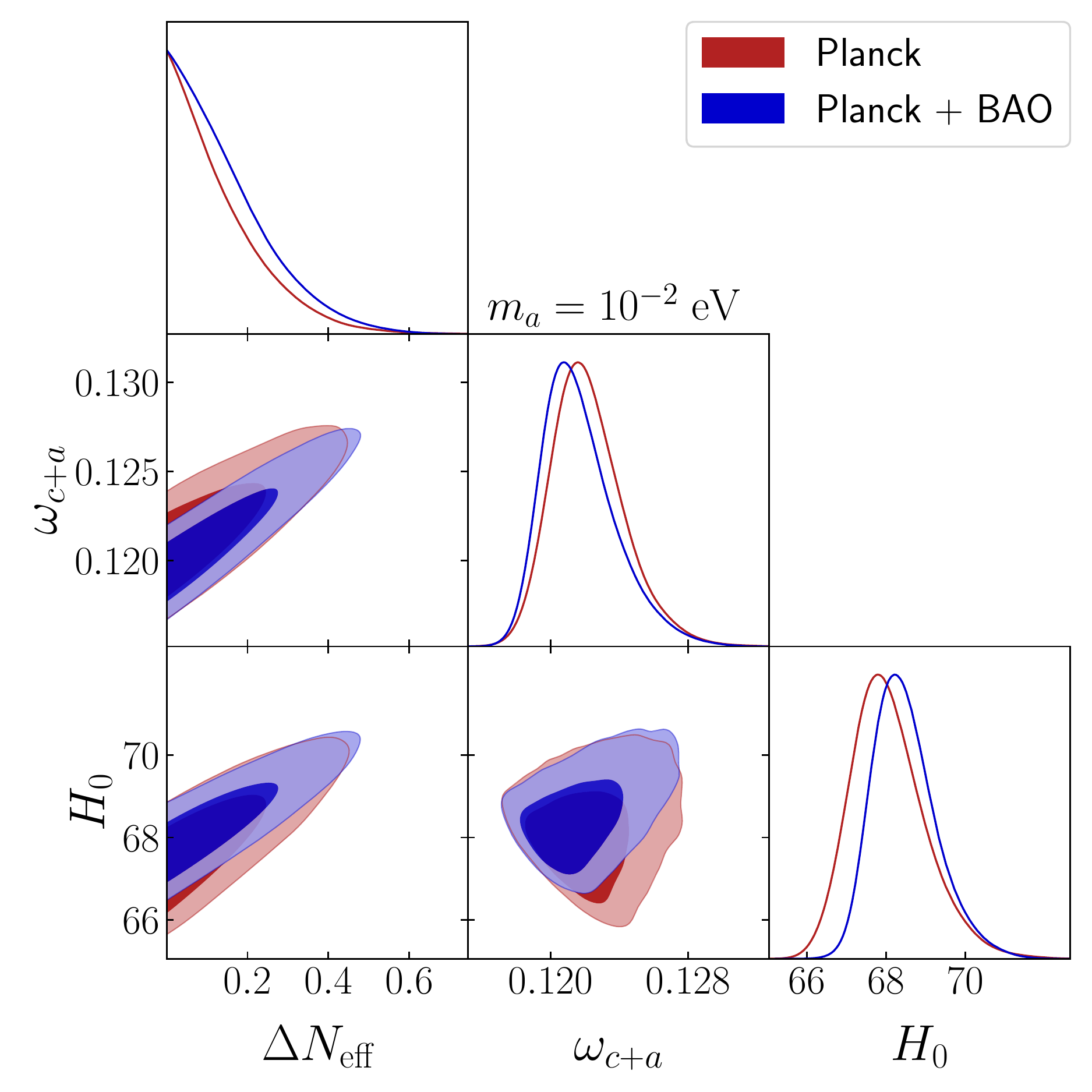}
	\includegraphics[width=0.45\textwidth]{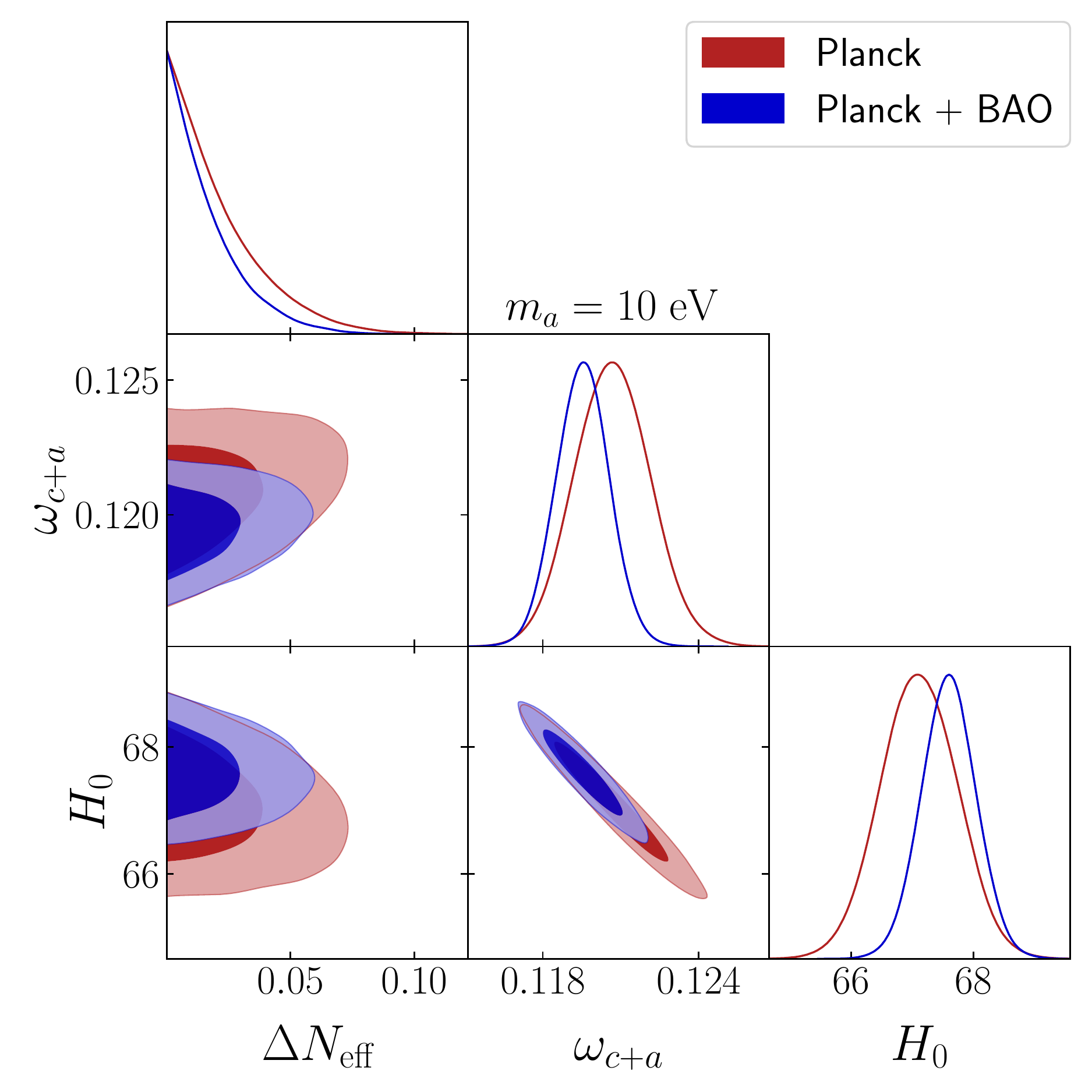}
	\caption{\textit{Left panel:} Triangle plot including 2D and 1D posteriors for different cosmological parameters in the case where the axion mass is fixed as $m_a = 10^{-2} \, \mathrm{eV}$. Note that, somehow unexpectedly, the inclusion of BAO data leads to a broader 1D distribution of $\Delta N_\mathrm{eff}$, see text for details. \textit{Right panel:} Same as left panel, for an axion mass $m_a = 10\,$eV. The inclusion of BAO data leads to tighter constraints on all parameters, including $\Delta N_\mathrm{eff}$.}
	\label{fig:Triangle-ma}
\end{figure}

We now turn to the discussion of the series of runs for the $\Lambda$CDM+$\Delta N_{\rm eff}$ model, with the axion mass fixed to a specific value as discussed in Sec.~\ref{sec:analysis}. Each run leads to a constraint on $\Delta N_{\rm eff}$ and on the parameters of $\Lambda$CDM, as shown in Fig.~\ref{fig:Triangle-ma} for the specific case $m_a = 10^{-2}\,$eV (left panel) and $m_a = 10\,$eV (right panel). The inclusion of BAO data over these two different runs acts in opposite directions with respect to constraining $\Delta N_{\rm eff}$: for lighter axions, the inclusion of BAO data leads to a broader 1D distribution of $\Delta N_\mathrm{eff}$, while for heavier axions the inclusion of BAO data leads to tighter constraints on all parameters, including $\Delta N_\mathrm{eff}$. This behaviour is ultimately due to the different role of the axion in the two regimes. Light axions are hot DM, so that their properties are mainly constrained by a measurement of $\Delta N_{\rm eff}$, while heavier axions behave as CDM at recombination and receive a tighter constraint from BAO data.

More in depth, for small values of $m_a$ the axion is a hot DM component which greatly contribute $N_{\rm eff}$ while not affecting $\omega_a$. However, the increase in $N_{\rm eff}$ leads to an increased $\omega_c$, as shown from the correlation between the two quantities in the left panel of Fig.~\ref{fig:Triangle-ma}. As CMB temperature and polarization data pin down the angular scale of sound horizon $\theta_s$, an increase in $N_{\rm eff}$ also leads to a higher $H_0$, as mentioned in the previous subsection. On the other hand, a massive axion would actively contribute to the CDM budget while not affecting $N_{\rm eff}$, see the right panel of Fig.~\ref{fig:Triangle-ma}. In this latter case, we obtain the bounds over $\Lambda$CDM although the total dark matter budget is given by the contribution $\omega_c + \omega_a$. This is a modification of late-time cosmology that does not alter the numerator of $\theta_s$ but it changes the angular diameter distance at the denominator, so that an increase in the CDM budget is compensated by a decrease in $H_0$. In fact, the correlation between these quantities in the right panel of Fig.~\ref{fig:Triangle-ma} is negative.

\begin{figure}
	\centering
	\includegraphics[width=0.7\textwidth]{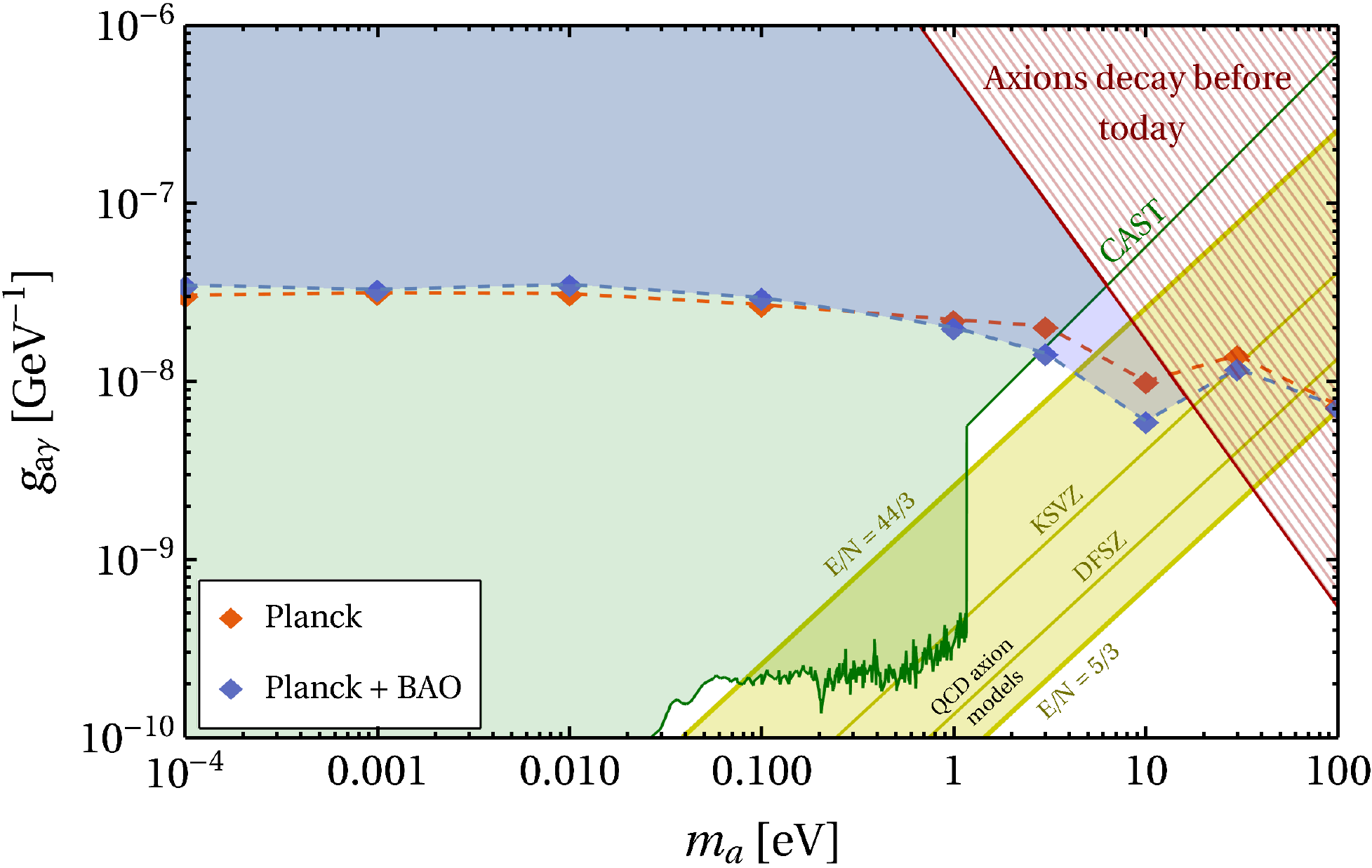}
	\caption{Summary of constraints in the axion-photon coupling $g_{a\gamma}$- axion mass $m_a$ plane. Colored diamonds connected with dashed lines are 95\% CL upper bounds on the axion-photon coupling $g_{a\gamma}$ as a function of the axion mass $m_a$, from {\it Planck} 2018 alone (orange) and {\it Planck} 2018+BAO (blue). The yellow band represents the representative QCD axion region considered (see text for details), including the KSVZ ($E/N = 0$) and the DFSZ ($E/N = 8/3$) models. The red hatched area labeled ``Axions decay before today'' indicates the region of parameter space where the axion lifetime before its decay into photons is smaller than the age of the Universe, see Eq.~\eqref{eq:axion-decay}. The green shaded area represents the region excluded by the CAST helioscope~\cite{CAST:2017uph,Irastorza:2018dyq}, with data taken from Ref.~\cite{ciaran_o_hare_2020_3932430}.}
	\label{fig:Constraints-gag}
\end{figure}

For a fixed mass, we convert the bound on $\Delta N_{\rm eff}$ into a bound on $g_{a\gamma}$ using Eq.~\eqref{eq:axion-photon-coupling}. The resulting upper bounds at 95\% CL are shown in Fig.~\ref{fig:Constraints-gag}, both from using the {\it Planck} 2018 dataset alone (orange diamonds connected with a dashed line) and in combination with BAO data (blue diamonds connected with a dashed line). 

The red hatched area (top right corner) excludes the region of the parameter space for which the decay rate of the axion into two photons is faster than the expansion rate of the Universe, so that the condition in Eq.~\eqref{eq:axion-decay} is not satisfied. Finally, the green region is excluded by the constraints on the axion-photon coupling derived by the CAST helioscope at CERN~\cite{CAST:2017uph}. The allowed region of the parameter space is what is left from the combination of the constraints above, i.e., the region in the bottom right part of the figure, on the left of the green curve, below the blue diamonds and on the right of the hatched region.

The yellow band in the figure identifies the region of the parameter space in which the axion-photon coupling is computed within a QCD axion theory and there is a solution of the QCD axion expected within grand unified models as (see e.g.\ Ref.~\cite{DiLuzio:2020wdo})
\begin{equation}
    g_{a\gamma}^0 = \frac{\alpha_{\rm EM}}{2\pi f_a} \frac{E}{N} \, ,
\end{equation}
where $E/N$ is the ratio between the colour and the electromagnetic axion anomalies. Generally speaking, different QCD axion theories predict different values of the $E/N$ ratio, which is expected to span from 5/3 to 44/3~\cite{Kim:1998va, DiLuzio:2016sbl, DiLuzio:2017pfr}. Explicit constructions that embed the axion into the SM include the DFSZ model~\cite{Dine:1981rt, Zhitnitsky:1980tq} in which a new Higgs doublet is introduced, and the KSVZ model~\cite{Kim:1979if, Shifman:1979if} in which exotic heavy quarks are introduced. The line marked ``KSVZ'' corresponds to the choice $E/N = 0$ which is obtained if the charge of the heavy quarks vanishes, while the line marked ``DFSZ'' corresponds to the choice $E/N = 8/3$. Note, that QCD axion solutions outside of this yellow band have also been constructed and are still viable~\cite{Farina:2016tgd, Agrawal:2017cmd, Hook:2018jle, DiLuzio:2021pxd, Sokolov:2021ydn}.

Similarly to what has been discussed for the results on $\Delta N_{\rm eff}$, the trend in the constraints with heavier axion masses can be understood as follows. If we increase the value of $m_a$ while keeping $g_{a\gamma}$ fixed, we are increasing the abundance of hot axions. This effect can be compensated by decreasing the value of $g_{a\gamma}$, which leads to an axion relic abundance that is diluted with respect to that of cosmic photons. This is the reason why the limit on $g_{a\gamma}$ tightens up when moving to higher values of $m_a$. For $m_a \gtrsim 10 \, \mathrm{eV}$, axions behave as CDM at recombination and the constraints on the parameter space result from requiring that the cosmological abundance of CDM does not exceed that of thermal axions. This leads to less stringent constraints on $g_{a\gamma}$, explaining the rise of the bound in this mass range. Finally, for $m_a \sim 100 \, \mathrm{eV}$ the constraints tightens again as the cold axion population is overproduced. The inclusion of BAO data weakens the constraints on $g_{a\gamma}$ for light axions of mass $m_a \lesssim 0.3 \, \mathrm{eV}$, compared to the case with {\it Planck} 2018 data alone. This occurs because for light masses, the axion relic abundance is negligible and the constraints are mainly due to the axion contribution to $\Delta N_\mathrm{eff}$. Since the inclusion of BAO data pushes $H_0$ towards higher values, see the posterior for $\Delta N_\mathrm{eff}$ in Fig.~\ref{fig:Triangle-ma}), the bound on $\Delta N_\mathrm{eff}$ and ultimately on $g_{a\gamma}$ relaxes. Instead, when higher values of the axion mass are considered ($m_a \gtrsim 0.3 \, \mathrm{eV}$), the combination of {\it Planck} 2018 and BAO datasets strengthens the constraints on the axion-photon coupling. In this case, the constraints are mainly due to the larger value of the axion relic abundance. Since the inclusion of BAO data improves the bound on $\omega_{c+a}$, the constraints on $\Delta N_\mathrm{eff}$ are also tightened.

The measurements of small-scale CMB anisotropies from ACT~\cite{ACT:2020frw,ACT:2020gnv} provide a further opportunity to constrain the properties of additional light relics. 
As reported in Ref.~\cite{ACT:2020frw}, combining ACT with \textit{Planck} data yields the bound $N_{\rm eff} = 2.74 \pm 0.17$ at 68\% CL, which signals a preference for a number of relativistic species smaller than the SM value 3.046.
Even though we leave a complete analysis including ACT data for future work, we anticipate how the bounds on the axion-photon coupling are improved in the region of parameter space where axions behave as dark radiation. Indeed, as we have already discussed, light axions are directly constrained by the measurement of $\Delta N_{\rm eff}$, thus we expect that they receive a higher benefit from the inclusion of ACT data with respect to heavier axions. Moreover, when dealing with heavier axions which behave as warm or cold DM, a more careful treatment of non-linearities is required. We plan to present a detailed analysis that includes ACT data for the full mass range explored in this work in a future release. In the case of an axion with mass $m_a = 10^{-4}$ eV, the bound on the axion-photon coupling from the combination of \textit{Planck}, BAO and ACT data reads $g_{a\gamma} < 2.19 \times 10^{-8} \; \mathrm{GeV}^{-1}$ at 95\% CL. This represents an improvement by a factor of about $1.6$ with respect to the case with only \textit{Planck}+BAO.

Finally, let us comment on how the cosmological constraints we have derived compare with those coming from laboratory experiments.\footnote{We do not consider the constraints from the axion haloscope ADMX since they hold for axion masses of order of few $\mu\mathrm{eV}$~\cite{ADMX:2021nhd}, which are smaller than those considered in this work. Moreover, haloscope experiments make the assumption that axions account for all the dark matter, while the constraints derived from cosmological observations do not have this restriction.} Our results are tighter than the bounds placed by the CAST helioscope for values of the axion mass $m_a \gtrsim 3 \, \mathrm{eV}$, according to Fig.~\ref{fig:Constraints-gag}. A conservative bound on $g_{a\gamma}$ for light axions has been derived based on the possible effect of these particles on the stellar evolution of massive stars, which translates into $g_{a\gamma}\lesssim 0.8\times 10^{-10}{\rm\,GeV^{-1}}$~\cite{Friedland:2012hj} which is tighter than what we obtain by using cosmological data. An even stronger bound arises from the comparison of the ratio of stars in horizontal over red giant branch found in a group of globular clusters with the corresponding predictions from accurate models of stellar evolution. This leads to $g_{a\gamma}\lesssim 6.6\times 10^{-11}{\rm\,GeV}^{-1}$ at 95\% CL~\cite{Ayala:2014pea}. Nevertheless, the bounds we derive are independent and complementary to laboratory searches and stellar evolution.

\begin{figure}
	\centering
	\includegraphics[width=0.8\textwidth]{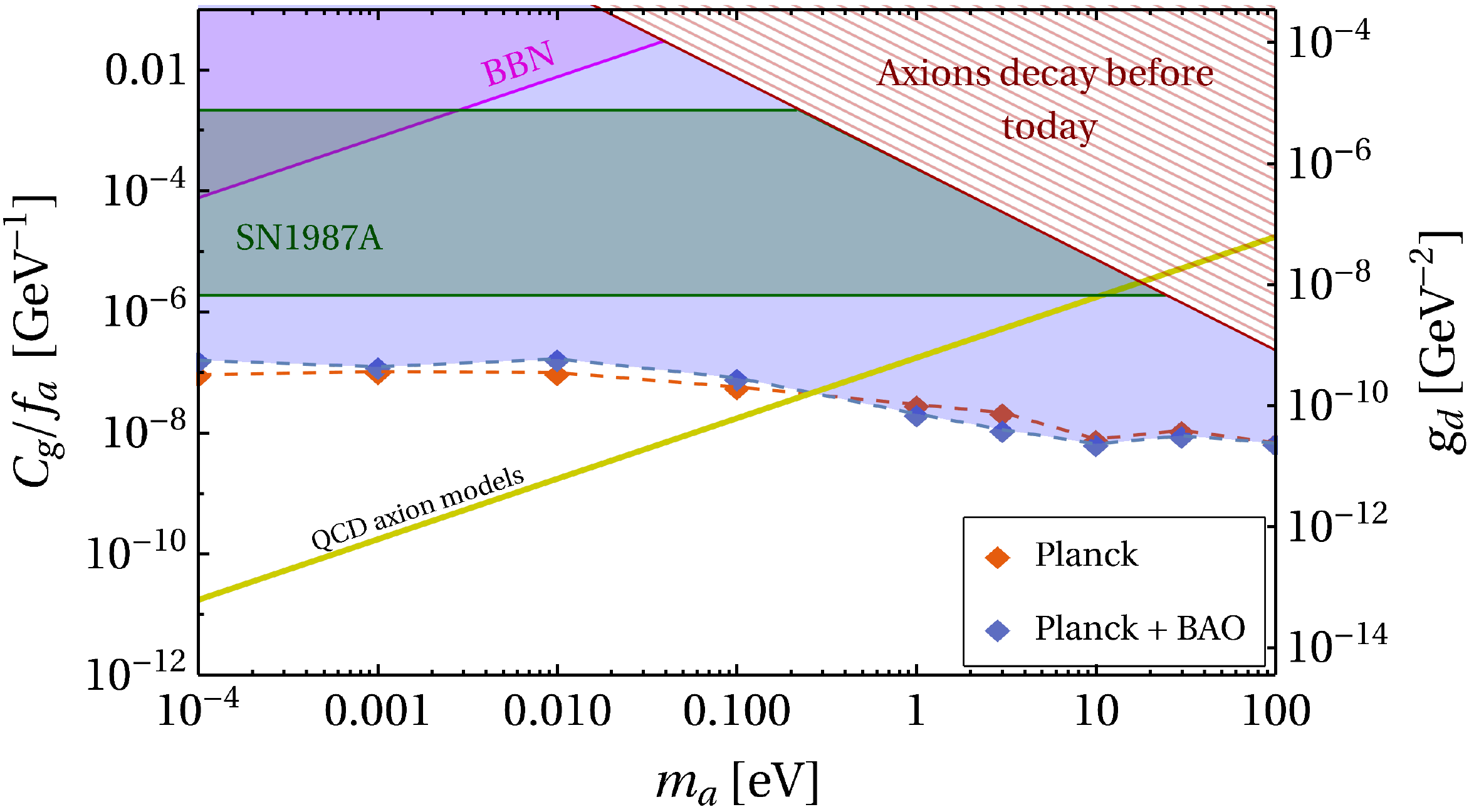}
	\caption{Summary of constraints in the axion-gluon coupling-axion mass plane. The vertical axis shows the axion-gluon coupling $C_g/f_a$ (left) or the EDM coupling $g_d$ (right) according to Eq.~\eqref{eq:coupling-gd}. Colored diamonds connected with dashed lines are 95\% CL upper bounds on the axion-gluon coupling as a function of the axion mass $m_a$, from {\it Planck} alone (orange) and {\it Planck}+BAO (blue). The yellow band represents the QCD axion region, which is determined by the uncertainties in the computation of the axion mass in Eq.~\eqref{eq:effective-mass} with $m_0 = 0$ and $C_g = 1$. The green shaded area represents the region of parameter space excluded by SN1987A energy loss consideration~\cite{Lucente:2022vuo}, as expressed in Eq.~\eqref{eq:gd-SN}. The red hatched area labeled ``Axions decay before today'' indicates the region of parameter space where the axion lifetime before its decay into photons is smaller than the age of the Universe, see Eq.~\eqref{eq:axion-decay}. The magenta shaded area is excluded by BBN considerations~\cite{Blum:2014vsa}. Notice that the latter constraints are derived under the assumption that axions account for the entirety of the DM, whereas the other bounds do not have this restriction. We stress that the cosmological bounds derived in this work directly constrain $C_g/f_a$. On the other hand, the results from SN1987A considerations translate into a bound on $g_d$, so that the conversion into a bound on $C_g/f_a$ is model-dependent via Eq.~\eqref{eq:coupling-gd}, see main text for details.}
	\label{fig:Constraints-gd}
\end{figure}

We now proceed to consider the bounds obtained on the axion-gluon coupling $C_g/f_a$. To derive these bounds, we use the results for the set of runs $\Lambda$CDM+$\Delta N_{\rm eff}$ for fixed axion mass to then convert the bounds over $\Delta N_{\rm eff}$ into a bound over $C_g/f_a$ using the method outlined in Sec.~\ref{sec:results_ma_DNeff}. In Fig.~\ref{fig:Constraints-gd} we show the 95\% CL upper bounds on the coupling $C_g/f_a$ (vertical left axis) in the case where the leading interaction in establishing a thermal population of axions is the axion-gluon interaction. The yellow band in the figure represents the region of parameter space in which the representative QCD axion models lie. Similarly to what has been discussed in relation to Fig.~\ref{fig:Constraints-gag}, the region above the blue (orange) diamonds is excluded by \textit{Planck} alone (\textit{Planck} in combination with BAO). These results are translated into a constraint on the coupling $g_d$ (vertical right axis) according to Eq.~\eqref{eq:coupling-gd}. Note, that the relation between $C_g$ and $g_d$ in Eq.~\eqref{eq:coupling-gd} is model-dependent, since it only takes into account the irreducible component of the neutron EDM coming from axion-gluon interaction.

Our results are compared with the constraints obtained from the considerations over the energy loss by astrophysical objects, in particular SN1987A excludes~\cite{Lucente:2022vuo} (see also Ref.~\cite{Graham:2013gfa} for a first attempt at estimating the lower bound of the excluded region)
\begin{equation}
    \label{eq:gd-SN}
    6.7\times 10^{-9}\,{\rm\,GeV^{-2}} \lesssim g_d \lesssim 7.7\times 10^{-6}\,{\rm\,GeV^{-2}} \, .
\end{equation}
This converts into the region $1.9\times 10^{-6}\,{\rm\,GeV^{-1}} \lesssim C_g/f_a \lesssim 2.2\times 10^{-3}\,{\rm\,GeV^{-1}}$ using Eq.~\eqref{eq:coupling-gd}, thus excluding the green horizontal region in Fig.~\ref{fig:Constraints-gd} labeled ``SN1987A''. Remarkably, the cosmological constraints derived in this work are stronger than already existing constraints derived from SN1987A energy loss considerations~\cite{Raffelt:1996wa}, as reported in Eq.~\eqref{eq:gd-SN}. For this reason, the results we obtain are the most stringent ones on the axion-gluon coupling in the mass range $10^{-4} \lesssim m_a/{\rm eV} \lesssim 100$. In analogy with the results in Fig.~\ref{fig:Constraints-gag}, the red hatched area in Fig.~\ref{fig:Constraints-gd} represents the region of parameter space where the lifetime of the axion before it decays into photons is smaller than the age of the Universe. The decay is due to the effective axion-photon coupling induced by the coupling with gluons, which is obtained by setting $g_{a\gamma}^0=0$ in Eq.~\eqref{eq:effective-coupling}.
Also shown are the bounds obtained from BBN considerations~\cite{Blum:2014vsa}, which we stress are obtained under the assumption that axions saturate the dark matter abundance. Instead, the same assumption is not enforced when deriving the results of this work.

We briefly comment on the implications of our analysis for the KSVZ axion. A proper derivation of the bounds for this case would require enforcing the relation between the axion mass and coupling as a prior in the Monte Carlo run, but this is beyond the scope of the paper. We can anyhow estimate the constraint we would get on $f_a$ (since $C_g=1$ for the QCD axion) from such an analysis by looking at the intersection of our bounds with the QCD axion region in Fig.~\ref{fig:Constraints-gd}. In this way we get $f_a \gtrsim 2\times 10^7\,$GeV, which is reasonably close to the value $f_a > 2.02 \times 10^7\,$GeV at 95\% CL quoted in Ref.~\cite{DEramo:2022nvb}.

Finally, similarly to the case of the axion-photon coupling, we briefly comment on how the bounds on the axion-gluon are improved with the inclusion of ACT data. For $m_a = 10^{-4}\,$eV, we find $C_g/f_a < 2.79 \times 10^{-8} \; \mathrm{GeV}^{-1}$ or, in terms of the EDM coupling, $g_d < 9.89 \times 10^{-11} \; \mathrm{GeV}^{-2}$ at 95\% CL, from \textit{Planck}+ACT+BAO. This represents an improvement by a factor of about $5.7$ with respect to the case with only \textit{Planck}+BAO. 

\section{Conclusions}
\label{sec:conclusions}

In this paper we have derived novel cosmological bounds on thermally-produced axion-like particles using cosmological data from {\it Planck} 2018 in combination with BAO measurements. We have explored two main mechanisms for producing the population of thermal axions in the early Universe, namely scattering via axion-gluon interactions or mediated by axion-photon interactions. For a given axion mass, the bounds on the effective number of species $N_{\rm eff}$ that are relativistic at recombination are mapped onto bounds on either the axion-gluon or the axion-photon coupling, depending on the model considered. In the case of the axion-gluon coupling, the mapping relies on the most updated results for the production rate of axions reported in Refs.~\cite{DEramo:2021psx,DEramo:2021lgb} and based on the results in Ref.~\cite{DiLuzio:2021vjd}.

On the model-building side, we have considered a Lagrangian for an axion-like particle (here ``axion'') which, in the chiral representation, leads to the effective axion mass and coupling to the photon in Eqs.~\eqref{eq:effective-mass} and~\eqref{eq:effective-coupling}, respectively. These expressions motivate the treatment of the axion mass and its interactions with SM particles as independent quantities and have not been implemented in previous literature. On the technical side, the novelty of our approach concerns the modified \texttt{CAMB} Boltzmann solver which has been adopted for the derivation of the results. More precisely, the version of \texttt{CAMB} used here propagates the axion field as a bosonic field, improving over previous approaches in which the axion had been treated as an additional degree of freedom that modifies the value of $N_{\rm eff}$ as in Eq.~\eqref{eq:Delta-Neff}.

We first consider a $\Lambda$CDM model augmented with the parameters $\Delta N_{\rm eff}$ and $m_a$, which are allowed to vary over the range in Table~\ref{tab:priors}. The constraints against {\it Planck} 2018 + BAO data lead to the bounds in Table~\ref{tab:constraints}, namely the axion mass lies in the range $m_a < 3.14\,$eV at 95\% CL; when the production is dominated by scatterings with photons, the axion-photon coupling is constrained as $g_{a\gamma} < 3.16 \times 10^{-8} \; \mathrm{GeV}^{-1}$ at 95\% CL, while if the production is dominated by scatterings with gluons we set the bound on the coupling in Eq.~\eqref{eq:coupling-gd} as $C_g/f_a \lesssim 1.1 \times 10^{-7}\; \mathrm{GeV}^{-1}$ or, equivalently, $g_d < 3.77 \times 10^{-10} \; \mathrm{GeV}^{-2}$ at 95\% CL. Note, that the bound on $g_d$ we derive is stronger by about one order of magnitude compared with what is obtained from energy loss considerations during SN1987A in Eq.~\eqref{eq:gd-SN}.

We have also considered a $\Lambda$CDM model in which only $\Delta N_{\rm eff}$ is allowed as an extra free parameter and the axion mass is fixed to a specific value as discussed in Sec.~\ref{sec:analysis}. For each fixed value of $m_a$, we scan the parameter space to derive the bound on $\Delta N_{\rm eff}$; we then collect these bounds to build a constrain on $\Delta N_{\rm eff}$ as a function of the axion mass, which is converted into a bound on the coupling of the axion with the specific SM particles considered. For the axion-photon coupling, this is summarized in Fig.~\ref{fig:Constraints-gag}. For small values of the axion mass ($m_a \lesssim 0.3 \, \mathrm{eV}$) the constraints derived are mainly due to the axion contribution to the effective number of relativistic species $\Delta N_{\rm eff}$. In this case, since the inclusion of BAO data pushes the Hubble constant $H_0$ to higher values, we find that the inclusion of BAO slightly weakens the bounds on the axion coupling to photons and gluons. On the other hand, for higher values of the axion mass ($m_a \gtrsim 0.3 \, \mathrm{eV}$) the constraints are mostly due to the axion contribution to the abundance of cold dark matter. Since this is better constrained when BAO measurements are taken into account, the inclusion of BAO data improves the bounds on the axion couplings. Remarkably, the cosmological constraints derived in this work are stronger than those obtained from the CAST helioscope at CERN for axion masses $m_a \gtrsim 3 \, \mathrm{eV}$. Similarly, the bounds on the axion-gluon coupling are reported in Fig.~\ref{fig:Constraints-gd}, placing a bound on the coupling $g_d$ which is about one order of magnitude stronger than what obtained by using energy loss considerations for the supernova event SN1987A~\cite{Raffelt:1996wa}.

Future surveys will test scenarios of exotic thermal relics with improved sensitivity, as reported by the CMB-S4 Collaboration which forecasts the constraint on $\Delta N_{\rm eff}^{\rm CMB} \lesssim 0.06$ at 95\% CL~\cite{CMB-S4:2016ple, Abazajian:2019eic}. This will allow to probe the existence for an axion with couplings to photons as weak as $g_{a\gamma}\simeq 1.07 \times 10^{-8} \, {\rm GeV}^{-1}$, or with couplings to gluons as weak as $C_g/f_a \simeq 8.3 \times 10^{-8}\; \mathrm{GeV}^{-1}$ or, equivalently, $g_d\simeq 2.93 \times 10^{-11} \, {\rm GeV}^{-2}$ for the case of very light axions. While the axion-photon coupling would not improve much, and would be weaker than the other existing bounds discussed in Fig.~\ref{fig:Constraints-gag}, the bound on the axion-gluon coupling would improve our current estimates by about one order of magnitude. Likewise, the whole range of masses could be discussed in light of these future results starting from an analysis similar to what has been demonstrated here.


\acknowledgments
We thank Francesco D'Eramo for comments on a preliminary version of the manuscript and for fruitful discussions. We are also grateful to Pierluca Carenza, Maurizio Giannotti, Giuseppe Lucente, Alessandro Mirizzi and Alessio Notari for valuable discussions. LC, MG and ML acknowledge support from the COSMOS network (www.cosmosnet.it) through the ASI (Italian Space Agency) Grants no.\ 2016-24-H.0, 2016-24-H.1-2018, and 2019-9-HH.0. We acknowledge the use of CINECA HPC resources from the InDark project in the framework of the INFN-CINECA agreement.

\bibliographystyle{JHEP}
\bibliography{bibliography.bib}

\providecommand{\href}[2]{#2}\begingroup\raggedright\begin{thebibliography}{100}

\bibitem{Peccei:1977hh}
R.~D. Peccei and H.~R. Quinn, \emph{{CP Conservation in the Presence of
  Instantons}},
  \href{http://dx.doi.org/10.1103/PhysRevLett.38.1440}{\emph{Phys. Rev. Lett.}
  {\bf 38} (1977) 1440--1443}.

\bibitem{Peccei:1977ur}
R.~D. Peccei and H.~R. Quinn, \emph{{Constraints Imposed by CP Conservation in
  the Presence of Instantons}},
  \href{http://dx.doi.org/10.1103/PhysRevD.16.1791}{\emph{Phys. Rev. D} {\bf
  16} (1977) 1791--1797}.

\bibitem{Weinberg:1977ma}
S.~Weinberg, \emph{{A New Light Boson?}},
  \href{http://dx.doi.org/10.1103/PhysRevLett.40.223}{\emph{Phys. Rev. Lett.}
  {\bf 40} (1978) 223--226}.

\bibitem{Wilczek:1977pj}
F.~Wilczek, \emph{{Problem of Strong $P$ and $T$ Invariance in the Presence of
  Instantons}}, \href{http://dx.doi.org/10.1103/PhysRevLett.40.279}{\emph{Phys.
  Rev. Lett.} {\bf 40} (1978) 279--282}.

\bibitem{Goldman:1977en}
J.~T. Goldman and C.~M. Hoffman, \emph{{Will the Axion Be Found Soon?}},
  \href{http://dx.doi.org/10.1103/PhysRevLett.40.220}{\emph{Phys. Rev. Lett.}
  {\bf 40} (1978) 220}.

\bibitem{Graham:2015ouw}
P.~W. Graham, I.~G. Irastorza, S.~K. Lamoreaux, A.~Lindner and K.~A. van
  Bibber, \emph{{Experimental Searches for the Axion and Axion-Like
  Particles}},
  \href{http://dx.doi.org/10.1146/annurev-nucl-102014-022120}{\emph{Ann. Rev.
  Nucl. Part. Sci.} {\bf 65} (2015) 485--514},
  [\href{http://arxiv.org/abs/1602.00039}{{\tt 1602.00039}}].

\bibitem{Irastorza:2018dyq}
I.~G. Irastorza and J.~Redondo, \emph{{New experimental approaches in the
  search for axion-like particles}},
  \href{http://dx.doi.org/10.1016/j.ppnp.2018.05.003}{\emph{Prog. Part. Nucl.
  Phys.} {\bf 102} (2018) 89--159},
  [\href{http://arxiv.org/abs/1801.08127}{{\tt 1801.08127}}].

\bibitem{Sikivie:2020zpn}
P.~Sikivie, \emph{{Invisible Axion Search Methods}},
  \href{http://dx.doi.org/10.1103/RevModPhys.93.015004}{\emph{Rev. Mod. Phys.}
  {\bf 93} (2021) 015004}, [\href{http://arxiv.org/abs/2003.02206}{{\tt
  2003.02206}}].

\bibitem{Abbott:1982af}
L.~F. Abbott and P.~Sikivie, \emph{{A Cosmological Bound on the Invisible
  Axion}}, \href{http://dx.doi.org/10.1016/0370-2693(83)90638-X}{\emph{Phys.
  Lett. B} {\bf 120} (1983) 133--136}.

\bibitem{Dine:1982ah}
M.~Dine and W.~Fischler, \emph{{The Not So Harmless Axion}},
  \href{http://dx.doi.org/10.1016/0370-2693(83)90639-1}{\emph{Phys. Lett. B}
  {\bf 120} (1983) 137--141}.

\bibitem{Preskill:1982cy}
J.~Preskill, M.~B. Wise and F.~Wilczek, \emph{{Cosmology of the Invisible
  Axion}}, \href{http://dx.doi.org/10.1016/0370-2693(83)90637-8}{\emph{Phys.
  Lett. B} {\bf 120} (1983) 127--132}.

\bibitem{Vilenkin:1982ks}
A.~Vilenkin and A.~E. Everett, \emph{{Cosmic Strings and Domain Walls in Models
  with Goldstone and PseudoGoldstone Bosons}},
  \href{http://dx.doi.org/10.1103/PhysRevLett.48.1867}{\emph{Phys. Rev. Lett.}
  {\bf 48} (1982) 1867--1870}.

\bibitem{Huang:1985tt}
M.~C. Huang and P.~Sikivie, \emph{{The Structure of Axionic Domain Walls}},
  \href{http://dx.doi.org/10.1103/PhysRevD.32.1560}{\emph{Phys. Rev. D} {\bf
  32} (1985) 1560}.

\bibitem{Davis:1986xc}
R.~L. Davis, \emph{{Cosmic Axions from Cosmic Strings}},
  \href{http://dx.doi.org/10.1016/0370-2693(86)90300-X}{\emph{Phys. Lett. B}
  {\bf 180} (1986) 225--230}.

\bibitem{Turner:1986tb}
M.~S. Turner, \emph{{Thermal Production of Not SO Invisible Axions in the Early
  Universe}}, \href{http://dx.doi.org/10.1103/PhysRevLett.59.2489}{\emph{Phys.
  Rev. Lett.} {\bf 59} (1987) 2489}.

\bibitem{McDonald:1993ex}
J.~McDonald, \emph{{Gauge singlet scalars as cold dark matter}},
  \href{http://dx.doi.org/10.1103/PhysRevD.50.3637}{\emph{Phys. Rev. D} {\bf
  50} (1994) 3637--3649}, [\href{http://arxiv.org/abs/hep-ph/0702143}{{\tt
  hep-ph/0702143}}].

\bibitem{Burgess:2000yq}
C.~P. Burgess, M.~Pospelov and T.~ter Veldhuis, \emph{{The Minimal model of
  nonbaryonic dark matter: A Singlet scalar}},
  \href{http://dx.doi.org/10.1016/S0550-3213(01)00513-2}{\emph{Nucl. Phys. B}
  {\bf 619} (2001) 709--728}, [\href{http://arxiv.org/abs/hep-ph/0011335}{{\tt
  hep-ph/0011335}}].

\bibitem{He:2009yd}
X.-G. He, T.~Li, X.-Q. Li, J.~Tandean and H.-C. Tsai, \emph{{The Simplest
  Dark-Matter Model, CDMS II Results, and Higgs Detection at LHC}},
  \href{http://dx.doi.org/10.1016/j.physletb.2010.04.026}{\emph{Phys. Lett. B}
  {\bf 688} (2010) 332--336}, [\href{http://arxiv.org/abs/0912.4722}{{\tt
  0912.4722}}].

\bibitem{Cacciapaglia:2019bqz}
G.~Cacciapaglia, G.~Ferretti, T.~Flacke and H.~Ser\^odio, \emph{{Light scalars
  in composite Higgs models}},
  \href{http://dx.doi.org/10.3389/fphy.2019.00022}{\emph{Front. in Phys.} {\bf
  7} (2019) 22}, [\href{http://arxiv.org/abs/1902.06890}{{\tt 1902.06890}}].

\bibitem{Lin:2022xbu}
W.~Lin, L.~Visinelli, D.~Xu and T.~T. Yanagida, \emph{{Neutrino astronomy as a
  probe of physics beyond the Standard Model: decay of sub-MeV $B$-$L$ gauge
  boson dark matter}},  \href{http://arxiv.org/abs/2202.04496}{{\tt
  2202.04496}}.

\bibitem{Svrcek:2006yi}
P.~Svrcek and E.~Witten, \emph{{Axions In String Theory}},
  \href{http://dx.doi.org/10.1088/1126-6708/2006/06/051}{\emph{JHEP} {\bf 06}
  (2006) 051}, [\href{http://arxiv.org/abs/hep-th/0605206}{{\tt
  hep-th/0605206}}].

\bibitem{Arvanitaki:2009fg}
A.~Arvanitaki, S.~Dimopoulos, S.~Dubovsky, N.~Kaloper and J.~March-Russell,
  \emph{{String Axiverse}},
  \href{http://dx.doi.org/10.1103/PhysRevD.81.123530}{\emph{Phys. Rev. D} {\bf
  81} (2010) 123530}, [\href{http://arxiv.org/abs/0905.4720}{{\tt 0905.4720}}].

\bibitem{Harigaya:2013vja}
K.~Harigaya, M.~Ibe, K.~Schmitz and T.~T. Yanagida, \emph{{Peccei-Quinn
  symmetry from a gauged discrete R symmetry}},
  \href{http://dx.doi.org/10.1103/PhysRevD.88.075022}{\emph{Phys. Rev. D} {\bf
  88} (2013) 075022}, [\href{http://arxiv.org/abs/1308.1227}{{\tt 1308.1227}}].

\bibitem{DiLuzio:2017tjx}
L.~Di~Luzio, E.~Nardi and L.~Ubaldi, \emph{{Accidental Peccei-Quinn symmetry
  protected to arbitrary order}},
  \href{http://dx.doi.org/10.1103/PhysRevLett.119.011801}{\emph{Phys. Rev.
  Lett.} {\bf 119} (2017) 011801}, [\href{http://arxiv.org/abs/1704.01122}{{\tt
  1704.01122}}].

\bibitem{Arias:2012az}
P.~Arias, D.~Cadamuro, M.~Goodsell, J.~Jaeckel, J.~Redondo and A.~Ringwald,
  \emph{{WISPy Cold Dark Matter}},
  \href{http://dx.doi.org/10.1088/1475-7516/2012/06/013}{\emph{JCAP} {\bf 06}
  (2012) 013}, [\href{http://arxiv.org/abs/1201.5902}{{\tt 1201.5902}}].

\bibitem{Visinelli:2017imh}
L.~Visinelli, \emph{{Light axion-like dark matter must be present during
  inflation}}, \href{http://dx.doi.org/10.1103/PhysRevD.96.023013}{\emph{Phys.
  Rev. D} {\bf 96} (2017) 023013}, [\href{http://arxiv.org/abs/1703.08798}{{\tt
  1703.08798}}].

\bibitem{Masso:2002np}
E.~Masso, F.~Rota and G.~Zsembinszki, \emph{{On axion thermalization in the
  early universe}},
  \href{http://dx.doi.org/10.1103/PhysRevD.66.023004}{\emph{Phys. Rev. D} {\bf
  66} (2002) 023004}, [\href{http://arxiv.org/abs/hep-ph/0203221}{{\tt
  hep-ph/0203221}}].

\bibitem{Graf:2010tv}
P.~Graf and F.~D. Steffen, \emph{{Thermal axion production in the primordial
  quark-gluon plasma}},
  \href{http://dx.doi.org/10.1103/PhysRevD.83.075011}{\emph{Phys. Rev. D} {\bf
  83} (2011) 075011}, [\href{http://arxiv.org/abs/1008.4528}{{\tt 1008.4528}}].

\bibitem{Salvio:2013iaa}
A.~Salvio, A.~Strumia and W.~Xue, \emph{{Thermal axion production}},
  \href{http://dx.doi.org/10.1088/1475-7516/2014/01/011}{\emph{JCAP} {\bf 01}
  (2014) 011}, [\href{http://arxiv.org/abs/1310.6982}{{\tt 1310.6982}}].

\bibitem{Berezhiani:1992rk}
Z.~G. Berezhiani, A.~S. Sakharov and M.~Y. Khlopov, \emph{{Primordial
  background of cosmological axions}}, {\emph{Sov. J. Nucl. Phys.} {\bf 55}
  (1992) 1063--1071}.

\bibitem{Chang:1993gm}
S.~Chang and K.~Choi, \emph{{Hadronic axion window and the big bang
  nucleosynthesis}},
  \href{http://dx.doi.org/10.1016/0370-2693(93)90656-3}{\emph{Phys. Lett. B}
  {\bf 316} (1993) 51--56}, [\href{http://arxiv.org/abs/hep-ph/9306216}{{\tt
  hep-ph/9306216}}].

\bibitem{Hannestad:2005df}
S.~Hannestad, A.~Mirizzi and G.~Raffelt, \emph{{New cosmological mass limit on
  thermal relic axions}},
  \href{http://dx.doi.org/10.1088/1475-7516/2005/07/002}{\emph{JCAP} {\bf 07}
  (2005) 002}, [\href{http://arxiv.org/abs/hep-ph/0504059}{{\tt
  hep-ph/0504059}}].

\bibitem{DEramo:2014urw}
F.~D'Eramo, L.~J. Hall and D.~Pappadopulo, \emph{{Multiverse Dark Matter: SUSY
  or Axions}}, \href{http://dx.doi.org/10.1007/JHEP11(2014)108}{\emph{JHEP}
  {\bf 11} (2014) 108}, [\href{http://arxiv.org/abs/1409.5123}{{\tt
  1409.5123}}].

\bibitem{Kawasaki:2015ofa}
M.~Kawasaki, M.~Yamada and T.~T. Yanagida, \emph{{Observable dark radiation
  from a cosmologically safe QCD axion}},
  \href{http://dx.doi.org/10.1103/PhysRevD.91.125018}{\emph{Phys. Rev. D} {\bf
  91} (2015) 125018}, [\href{http://arxiv.org/abs/1504.04126}{{\tt
  1504.04126}}].

\bibitem{Ferreira:2020bpb}
R.~Z. Ferreira, A.~Notari and F.~Rompineve,
  \emph{{Dine-Fischler-Srednicki-Zhitnitsky axion in the CMB}},
  \href{http://dx.doi.org/10.1103/PhysRevD.103.063524}{\emph{Phys. Rev. D} {\bf
  103} (2021) 063524}, [\href{http://arxiv.org/abs/2012.06566}{{\tt
  2012.06566}}].

\bibitem{DEramo:2018vss}
F.~D'Eramo, R.~Z. Ferreira, A.~Notari and J.~L. Bernal, \emph{{Hot Axions and
  the $H_0$ tension}},
  \href{http://dx.doi.org/10.1088/1475-7516/2018/11/014}{\emph{JCAP} {\bf 11}
  (2018) 014}, [\href{http://arxiv.org/abs/1808.07430}{{\tt 1808.07430}}].

\bibitem{Ferreira:2018vjj}
R.~Z. Ferreira and A.~Notari, \emph{{Observable Windows for the QCD Axion
  Through the Number of Relativistic Species}},
  \href{http://dx.doi.org/10.1103/PhysRevLett.120.191301}{\emph{Phys. Rev.
  Lett.} {\bf 120} (2018) 191301}, [\href{http://arxiv.org/abs/1801.06090}{{\tt
  1801.06090}}].

\bibitem{Arias-Aragon:2020qtn}
F.~Arias-Arag\'on, F.~D'eramo, R.~Z. Ferreira, L.~Merlo and A.~Notari,
  \emph{{Cosmic Imprints of XENON1T Axions}},
  \href{http://dx.doi.org/10.1088/1475-7516/2020/11/025}{\emph{JCAP} {\bf 11}
  (2020) 025}, [\href{http://arxiv.org/abs/2007.06579}{{\tt 2007.06579}}].

\bibitem{Dror:2021nyr}
J.~A. Dror, H.~Murayama and N.~L. Rodd, \emph{{Cosmic axion background}},
  \href{http://dx.doi.org/10.1103/PhysRevD.103.115004}{\emph{Phys. Rev. D} {\bf
  103} (2021) 115004}, [\href{http://arxiv.org/abs/2101.09287}{{\tt
  2101.09287}}].

\bibitem{Bashinsky:2003tk}
S.~Bashinsky and U.~Seljak, \emph{{Neutrino perturbations in CMB anisotropy and
  matter clustering}},
  \href{http://dx.doi.org/10.1103/PhysRevD.69.083002}{\emph{Phys. Rev. D} {\bf
  69} (2004) 083002}, [\href{http://arxiv.org/abs/astro-ph/0310198}{{\tt
  astro-ph/0310198}}].

\bibitem{Hou:2011ec}
Z.~Hou, R.~Keisler, L.~Knox, M.~Millea and C.~Reichardt, \emph{{How Massless
  Neutrinos Affect the Cosmic Microwave Background Damping Tail}},
  \href{http://dx.doi.org/10.1103/PhysRevD.87.083008}{\emph{Phys. Rev. D} {\bf
  87} (2013) 083008}, [\href{http://arxiv.org/abs/1104.2333}{{\tt 1104.2333}}].

\bibitem{Weinberg:2013kea}
S.~Weinberg, \emph{{Goldstone Bosons as Fractional Cosmic Neutrinos}},
  \href{http://dx.doi.org/10.1103/PhysRevLett.110.241301}{\emph{Phys. Rev.
  Lett.} {\bf 110} (2013) 241301}, [\href{http://arxiv.org/abs/1305.1971}{{\tt
  1305.1971}}].

\bibitem{Mangano:2001iu}
G.~Mangano, G.~Miele, S.~Pastor and M.~Peloso, \emph{{A Precision calculation
  of the effective number of cosmological neutrinos}},
  \href{http://dx.doi.org/10.1016/S0370-2693(02)01622-2}{\emph{Phys. Lett. B}
  {\bf 534} (2002) 8--16}, [\href{http://arxiv.org/abs/astro-ph/0111408}{{\tt
  astro-ph/0111408}}].

\bibitem{deSalas:2016ztq}
P.~F. de~Salas and S.~Pastor, \emph{{Relic neutrino decoupling with flavour
  oscillations revisited}},
  \href{http://dx.doi.org/10.1088/1475-7516/2016/07/051}{\emph{JCAP} {\bf 07}
  (2016) 051}, [\href{http://arxiv.org/abs/1606.06986}{{\tt 1606.06986}}].

\bibitem{Akita:2020szl}
K.~Akita and M.~Yamaguchi, \emph{{A precision calculation of relic neutrino
  decoupling}},
  \href{http://dx.doi.org/10.1088/1475-7516/2020/08/012}{\emph{JCAP} {\bf 08}
  (2020) 012}, [\href{http://arxiv.org/abs/2005.07047}{{\tt 2005.07047}}].

\bibitem{Bennett:2020zkv}
J.~J. Bennett, G.~Buldgen, P.~F. De~Salas, M.~Drewes, S.~Gariazzo, S.~Pastor
  et~al., \emph{{Towards a precision calculation of $N_{\rm eff}$ in the
  Standard Model II: Neutrino decoupling in the presence of flavour
  oscillations and finite-temperature QED}},
  \href{http://dx.doi.org/10.1088/1475-7516/2021/04/073}{\emph{JCAP} {\bf 04}
  (2021) 073}, [\href{http://arxiv.org/abs/2012.02726}{{\tt 2012.02726}}].

\bibitem{Planck:2018vyg}
{\scshape Planck} collaboration, N.~Aghanim et~al., \emph{{Planck 2018 results.
  VI. Cosmological parameters}},
  \href{http://dx.doi.org/10.1051/0004-6361/201833910}{\emph{Astron.
  Astrophys.} {\bf 641} (2020) A6},
  [\href{http://arxiv.org/abs/1807.06209}{{\tt 1807.06209}}].

\bibitem{CMB-S4:2016ple}
{\scshape CMB-S4} collaboration, K.~N. Abazajian et~al., \emph{{CMB-S4 Science
  Book, First Edition}},  \href{http://arxiv.org/abs/1610.02743}{{\tt
  1610.02743}}.

\bibitem{Abazajian:2019eic}
K.~Abazajian et~al., \emph{{CMB-S4 Science Case, Reference Design, and Project
  Plan}},  \href{http://arxiv.org/abs/1907.04473}{{\tt 1907.04473}}.

\bibitem{Giare:2020vzo}
W.~Giar\`e, E.~Di~Valentino, A.~Melchiorri and O.~Mena, \emph{{New cosmological
  bounds on hot relics: axions and neutrinos}},
  \href{http://dx.doi.org/10.1093/mnras/stab1442}{\emph{Mon. Not. Roy. Astron.
  Soc.} {\bf 505} (2021) 2703--2711},
  [\href{http://arxiv.org/abs/2011.14704}{{\tt 2011.14704}}].

\bibitem{Hannestad:2003ye}
S.~Hannestad and G.~Raffelt, \emph{{Cosmological mass limits on neutrinos,
  axions, and other light particles}},
  \href{http://dx.doi.org/10.1088/1475-7516/2004/04/008}{\emph{JCAP} {\bf 04}
  (2004) 008}, [\href{http://arxiv.org/abs/hep-ph/0312154}{{\tt
  hep-ph/0312154}}].

\bibitem{Hannestad:2007dd}
S.~Hannestad, A.~Mirizzi, G.~G. Raffelt and Y.~Y.~Y. Wong, \emph{{Cosmological
  constraints on neutrino plus axion hot dark matter}},
  \href{http://dx.doi.org/10.1088/1475-7516/2007/08/015}{\emph{JCAP} {\bf 08}
  (2007) 015}, [\href{http://arxiv.org/abs/0706.4198}{{\tt 0706.4198}}].

\bibitem{Melchiorri:2007cd}
A.~Melchiorri, O.~Mena and A.~Slosar, \emph{{An improved cosmological bound on
  the thermal axion mass}},
  \href{http://dx.doi.org/10.1103/PhysRevD.76.041303}{\emph{Phys. Rev. D} {\bf
  76} (2007) 041303}, [\href{http://arxiv.org/abs/0705.2695}{{\tt 0705.2695}}].

\bibitem{Archidiacono:2013cha}
M.~Archidiacono, S.~Hannestad, A.~Mirizzi, G.~Raffelt and Y.~Y.~Y. Wong,
  \emph{{Axion hot dark matter bounds after Planck}},
  \href{http://dx.doi.org/10.1088/1475-7516/2013/10/020}{\emph{JCAP} {\bf 10}
  (2013) 020}, [\href{http://arxiv.org/abs/1307.0615}{{\tt 1307.0615}}].

\bibitem{Giusarma:2014zza}
E.~Giusarma, E.~Di~Valentino, M.~Lattanzi, A.~Melchiorri and O.~Mena,
  \emph{{Relic Neutrinos, thermal axions and cosmology in early 2014}},
  \href{http://dx.doi.org/10.1103/PhysRevD.90.043507}{\emph{Phys. Rev. D} {\bf
  90} (2014) 043507}, [\href{http://arxiv.org/abs/1403.4852}{{\tt 1403.4852}}].

\bibitem{DiValentino:2014zna}
E.~Di~Valentino, E.~Giusarma, M.~Lattanzi, A.~Melchiorri and O.~Mena,
  \emph{{Axion cold dark matter: status after Planck and BICEP2}},
  \href{http://dx.doi.org/10.1103/PhysRevD.90.043534}{\emph{Phys. Rev. D} {\bf
  90} (2014) 043534}, [\href{http://arxiv.org/abs/1405.1860}{{\tt 1405.1860}}].

\bibitem{DiValentino:2015wba}
E.~Di~Valentino, E.~Giusarma, M.~Lattanzi, O.~Mena, A.~Melchiorri and J.~Silk,
  \emph{{Cosmological Axion and neutrino mass constraints from Planck 2015
  temperature and polarization data}},
  \href{http://dx.doi.org/10.1016/j.physletb.2015.11.025}{\emph{Phys. Lett. B}
  {\bf 752} (2016) 182--185}, [\href{http://arxiv.org/abs/1507.08665}{{\tt
  1507.08665}}].

\bibitem{ACT:2020gnv}
{\scshape ACT} collaboration, S.~Aiola et~al., \emph{{The Atacama Cosmology
  Telescope: DR4 Maps and Cosmological Parameters}},
  \href{http://dx.doi.org/10.1088/1475-7516/2020/12/047}{\emph{JCAP} {\bf 12}
  (2020) 047}, [\href{http://arxiv.org/abs/2007.07288}{{\tt 2007.07288}}].

\bibitem{ACT:2020frw}
{\scshape ACT} collaboration, S.~K. Choi et~al., \emph{{The Atacama Cosmology
  Telescope: a measurement of the Cosmic Microwave Background power spectra at
  98 and 150 GHz}},
  \href{http://dx.doi.org/10.1088/1475-7516/2020/12/045}{\emph{JCAP} {\bf 12}
  (2020) 045}, [\href{http://arxiv.org/abs/2007.07289}{{\tt 2007.07289}}].

\bibitem{SPT-3G:2021wgf}
{\scshape SPT-3G} collaboration, L.~Balkenhol et~al., \emph{{Constraints on
  \ensuremath{\Lambda}CDM extensions from the SPT-3G 2018 EE and TE power
  spectra}}, \href{http://dx.doi.org/10.1103/PhysRevD.104.083509}{\emph{Phys.
  Rev. D} {\bf 104} (2021) 083509},
  [\href{http://arxiv.org/abs/2103.13618}{{\tt 2103.13618}}].

\bibitem{SPT-3G:2021eoc}
{\scshape SPT-3G} collaboration, D.~Dutcher et~al., \emph{{Measurements of the
  E-mode polarization and temperature-E-mode correlation of the CMB from SPT-3G
  2018 data}}, \href{http://dx.doi.org/10.1103/PhysRevD.104.022003}{\emph{Phys.
  Rev. D} {\bf 104} (2021) 022003},
  [\href{http://arxiv.org/abs/2101.01684}{{\tt 2101.01684}}].

\bibitem{DES:2022qpf}
{\scshape DES} collaboration, C.~Doux et~al., \emph{{Dark Energy Survey Year 3
  results: cosmological constraints from the analysis of cosmic shear in
  harmonic space}},  \href{http://arxiv.org/abs/2203.07128}{{\tt 2203.07128}}.

\bibitem{DES:2021wwk}
{\scshape DES} collaboration, T.~M.~C. Abbott et~al., \emph{{Dark Energy Survey
  Year 3 results: Cosmological constraints from galaxy clustering and weak
  lensing}}, \href{http://dx.doi.org/10.1103/PhysRevD.105.023520}{\emph{Phys.
  Rev. D} {\bf 105} (2022) 023520},
  [\href{http://arxiv.org/abs/2105.13549}{{\tt 2105.13549}}].

\bibitem{BOSS:2016wmc}
{\scshape BOSS} collaboration, S.~Alam et~al., \emph{{The clustering of
  galaxies in the completed SDSS-III Baryon Oscillation Spectroscopic Survey:
  cosmological analysis of the DR12 galaxy sample}},
  \href{http://dx.doi.org/10.1093/mnras/stx721}{\emph{Mon. Not. Roy. Astron.
  Soc.} {\bf 470} (2017) 2617--2652},
  [\href{http://arxiv.org/abs/1607.03155}{{\tt 1607.03155}}].

\bibitem{KiDS:2020suj}
{\scshape KiDS} collaboration, M.~Asgari et~al., \emph{{KiDS-1000 Cosmology:
  Cosmic shear constraints and comparison between two point statistics}},
  \href{http://dx.doi.org/10.1051/0004-6361/202039070}{\emph{Astron.
  Astrophys.} {\bf 645} (2021) A104},
  [\href{http://arxiv.org/abs/2007.15633}{{\tt 2007.15633}}].

\bibitem{KiDS:2020ghu}
{\scshape KiDS} collaboration, T.~Tr\"oster et~al., \emph{{KiDS-1000 Cosmology:
  Constraints beyond flat \ensuremath{\Lambda}CDM}},
  \href{http://dx.doi.org/10.1051/0004-6361/202039805}{\emph{Astron.
  Astrophys.} {\bf 649} (2021) A88},
  [\href{http://arxiv.org/abs/2010.16416}{{\tt 2010.16416}}].

\bibitem{Sgier:2021bzf}
R.~Sgier, C.~Lorenz, A.~Refregier, J.~Fluri, D.~Z\"urcher and F.~Tarsitano,
  \emph{{Combined $13\times2$-point analysis of the Cosmic Microwave Background
  and Large-Scale Structure: implications for the $S_8$-tension and neutrino
  mass constraints}},  \href{http://arxiv.org/abs/2110.03815}{{\tt
  2110.03815}}.

\bibitem{Carenza:2021ebx}
P.~Carenza, M.~Lattanzi, A.~Mirizzi and F.~Forastieri, \emph{{Thermal axions
  with multi-eV masses are possible in low-reheating scenarios}},
  \href{http://dx.doi.org/10.1088/1475-7516/2021/07/031}{\emph{JCAP} {\bf 07}
  (2021) 031}, [\href{http://arxiv.org/abs/2104.03982}{{\tt 2104.03982}}].

\bibitem{Marsh:2015xka}
D.~J.~E. Marsh, \emph{{Axion Cosmology}},
  \href{http://dx.doi.org/10.1016/j.physrep.2016.06.005}{\emph{Phys. Rept.}
  {\bf 643} (2016) 1--79}, [\href{http://arxiv.org/abs/1510.07633}{{\tt
  1510.07633}}].

\bibitem{DiLuzio:2020wdo}
L.~Di~Luzio, M.~Giannotti, E.~Nardi and L.~Visinelli, \emph{{The landscape of
  QCD axion models}},
  \href{http://dx.doi.org/10.1016/j.physrep.2020.06.002}{\emph{Phys. Rept.}
  {\bf 870} (2020) 1--117}, [\href{http://arxiv.org/abs/2003.01100}{{\tt
  2003.01100}}].

\bibitem{Crewther:1979pi}
R.~J. Crewther, P.~Di~Vecchia, G.~Veneziano and E.~Witten, \emph{{Chiral
  Estimate of the Electric Dipole Moment of the Neutron in Quantum
  Chromodynamics}},
  \href{http://dx.doi.org/10.1016/0370-2693(79)90128-X}{\emph{Phys. Lett. B}
  {\bf 88} (1979) 123}.

\bibitem{Pospelov:1999ha}
M.~Pospelov and A.~Ritz, \emph{{Theta induced electric dipole moment of the
  neutron via QCD sum rules}},
  \href{http://dx.doi.org/10.1103/PhysRevLett.83.2526}{\emph{Phys. Rev. Lett.}
  {\bf 83} (1999) 2526--2529}, [\href{http://arxiv.org/abs/hep-ph/9904483}{{\tt
  hep-ph/9904483}}].

\bibitem{Graham:2013gfa}
P.~W. Graham and S.~Rajendran, \emph{{New Observables for Direct Detection of
  Axion Dark Matter}},
  \href{http://dx.doi.org/10.1103/PhysRevD.88.035023}{\emph{Phys. Rev. D} {\bf
  88} (2013) 035023}, [\href{http://arxiv.org/abs/1306.6088}{{\tt 1306.6088}}].

\bibitem{Baumann:2016wac}
D.~Baumann, D.~Green and B.~Wallisch, \emph{{New Target for Cosmic Axion
  Searches}},
  \href{http://dx.doi.org/10.1103/PhysRevLett.117.171301}{\emph{Phys. Rev.
  Lett.} {\bf 117} (2016) 171301}, [\href{http://arxiv.org/abs/1604.08614}{{\tt
  1604.08614}}].

\bibitem{Georgi:1986df}
H.~Georgi, D.~B. Kaplan and L.~Randall, \emph{{Manifesting the Invisible Axion
  at Low-energies}},
  \href{http://dx.doi.org/10.1016/0370-2693(86)90688-X}{\emph{Phys. Lett. B}
  {\bf 169} (1986) 73--78}.

\bibitem{Aloni:2018vki}
D.~Aloni, Y.~Soreq and M.~Williams, \emph{{Coupling QCD-Scale Axionlike
  Particles to Gluons}},
  \href{http://dx.doi.org/10.1103/PhysRevLett.123.031803}{\emph{Phys. Rev.
  Lett.} {\bf 123} (2019) 031803}, [\href{http://arxiv.org/abs/1811.03474}{{\tt
  1811.03474}}].

\bibitem{10.1093/ptep/ptaa104}
{\scshape Particle Data Group} collaboration, P.~A. Zyla et~al., \emph{{Review
  of Particle Physics}},
  \href{http://dx.doi.org/10.1093/ptep/ptaa104}{\emph{Progress of Theoretical
  and Experimental Physics} {\bf 2020} (08, 2020) },
  [\href{http://arxiv.org/abs/https://academic.oup.com/ptep/article-pdf/2020/8/083C01/34673722/ptaa104.pdf}{{\tt
  https://academic.oup.com/ptep/article-pdf/2020/8/083C01/34673722/ptaa104.pdf}}].

\bibitem{Peccei:2006as}
R.~D. Peccei, \emph{{The Strong CP problem and axions}},
  \href{http://dx.doi.org/10.1007/978-3-540-73518-2_1}{\emph{Lect. Notes Phys.}
  {\bf 741} (2008) 3--17}, [\href{http://arxiv.org/abs/hep-ph/0607268}{{\tt
  hep-ph/0607268}}].

\bibitem{Blum:2014vsa}
K.~Blum, R.~T. D'Agnolo, M.~Lisanti and B.~R. Safdi, \emph{{Constraining Axion
  Dark Matter with Big Bang Nucleosynthesis}},
  \href{http://dx.doi.org/10.1016/j.physletb.2014.07.059}{\emph{Phys. Lett. B}
  {\bf 737} (2014) 30--33}, [\href{http://arxiv.org/abs/1401.6460}{{\tt
  1401.6460}}].

\bibitem{Saikawa:2018rcs}
K.~Saikawa and S.~Shirai, \emph{{Primordial gravitational waves, precisely: The
  role of thermodynamics in the Standard Model}},
  \href{http://dx.doi.org/10.1088/1475-7516/2018/05/035}{\emph{JCAP} {\bf 05}
  (2018) 035}, [\href{http://arxiv.org/abs/1803.01038}{{\tt 1803.01038}}].

\bibitem{Drees:2015exa}
M.~Drees, F.~Hajkarim and E.~R. Schmitz, \emph{{The Effects of QCD Equation of
  State on the Relic Density of WIMP Dark Matter}},
  \href{http://dx.doi.org/10.1088/1475-7516/2015/06/025}{\emph{JCAP} {\bf 06}
  (2015) 025}, [\href{http://arxiv.org/abs/1503.03513}{{\tt 1503.03513}}].

\bibitem{DEramo:2021lgb}
F.~D'Eramo, F.~Hajkarim and S.~Yun, \emph{{Thermal QCD Axions across
  Thresholds}}, \href{http://dx.doi.org/10.1007/JHEP10(2021)224}{\emph{JHEP}
  {\bf 10} (2021) 224}, [\href{http://arxiv.org/abs/2108.05371}{{\tt
  2108.05371}}].

\bibitem{Mather:1993ij}
J.~C. Mather et~al., \emph{{Measurement of the Cosmic Microwave Background
  spectrum by the COBE FIRAS instrument}},
  \href{http://dx.doi.org/10.1086/173574}{\emph{Astrophys. J.} {\bf 420} (1994)
  439--444}.

\bibitem{Fixsen:1996nj}
D.~J. Fixsen, E.~S. Cheng, J.~M. Gales, J.~C. Mather, R.~A. Shafer and E.~L.
  Wright, \emph{{The Cosmic Microwave Background spectrum from the full COBE
  FIRAS data set}}, \href{http://dx.doi.org/10.1086/178173}{\emph{Astrophys.
  J.} {\bf 473} (1996) 576}, [\href{http://arxiv.org/abs/astro-ph/9605054}{{\tt
  astro-ph/9605054}}].

\bibitem{Bennett:2019ewm}
J.~J. Bennett, G.~Buldgen, M.~Drewes and Y.~Y.~Y. Wong, \emph{{Towards a
  precision calculation of the effective number of neutrinos $N_{\rm eff}$ in
  the Standard Model I: the QED equation of state}},
  \href{http://dx.doi.org/10.1088/1475-7516/2020/03/003}{\emph{JCAP} {\bf 03}
  (2020) 003}, [\href{http://arxiv.org/abs/1911.04504}{{\tt 1911.04504}}].

\bibitem{Froustey:2020mcq}
J.~Froustey, C.~Pitrou and M.~C. Volpe, \emph{{Neutrino decoupling including
  flavour oscillations and primordial nucleosynthesis}},
  \href{http://dx.doi.org/10.1088/1475-7516/2020/12/015}{\emph{JCAP} {\bf 12}
  (2020) 015}, [\href{http://arxiv.org/abs/2008.01074}{{\tt 2008.01074}}].

\bibitem{Weldon:1990iw}
H.~A. Weldon, \emph{{Reformulation of finite temperature dilepton production}},
  \href{http://dx.doi.org/10.1103/PhysRevD.42.2384}{\emph{Phys. Rev. D} {\bf
  42} (1990) 2384--2387}.

\bibitem{Gale:1990pn}
C.~Gale and J.~I. Kapusta, \emph{{Vector dominance model at finite
  temperature}},
  \href{http://dx.doi.org/10.1016/0550-3213(91)90459-B}{\emph{Nucl. Phys. B}
  {\bf 357} (1991) 65--89}.

\bibitem{Raffelt:1996wa}
G.~G. Raffelt, \emph{{Stars as laboratories for fundamental physics}: {The
  astrophysics of neutrinos, axions, and other weakly interacting particles}}.
\newblock University of Chicago Press, 5, 1996.

\bibitem{DiLuzio:2021vjd}
L.~Di~Luzio, G.~Martinelli and G.~Piazza, \emph{{Breakdown of chiral
  perturbation theory for the axion hot dark matter bound}},
  \href{http://dx.doi.org/10.1103/PhysRevLett.126.241801}{\emph{Phys. Rev.
  Lett.} {\bf 126} (2021) 241801}, [\href{http://arxiv.org/abs/2101.10330}{{\tt
  2101.10330}}].

\bibitem{DEramo:2021psx}
F.~D'Eramo, F.~Hajkarim and S.~Yun, \emph{{Thermal Axion Production at Low
  Temperatures: A Smooth Treatment of the QCD Phase Transition}},
  \href{http://dx.doi.org/10.1103/PhysRevLett.128.152001}{\emph{Phys. Rev.
  Lett.} {\bf 128} (2022) 152001}, [\href{http://arxiv.org/abs/2108.04259}{{\tt
  2108.04259}}].

\bibitem{Bolz:2000fu}
M.~Bolz, A.~Brandenburg and W.~Buchmuller, \emph{{Thermal production of
  gravitinos}},
  \href{http://dx.doi.org/10.1016/S0550-3213(01)00132-8}{\emph{Nucl. Phys. B}
  {\bf 606} (2001) 518--544}, [\href{http://arxiv.org/abs/hep-ph/0012052}{{\tt
  hep-ph/0012052}}].

\bibitem{Cadamuro:2011fd}
D.~Cadamuro and J.~Redondo, \emph{{Cosmological bounds on pseudo
  Nambu-Goldstone bosons}},
  \href{http://dx.doi.org/10.1088/1475-7516/2012/02/032}{\emph{JCAP} {\bf 02}
  (2012) 032}, [\href{http://arxiv.org/abs/1110.2895}{{\tt 1110.2895}}].

\bibitem{Millea:2020xxp}
M.~Millea, \emph{{New cosmological bounds on axions in the XENON1T window}},
  \href{http://arxiv.org/abs/2007.05659}{{\tt 2007.05659}}.

\bibitem{HotQCD:2014kol}
{\scshape HotQCD} collaboration, A.~Bazavov et~al., \emph{{Equation of state in
  ( 2+1 )-flavor QCD}},
  \href{http://dx.doi.org/10.1103/PhysRevD.90.094503}{\emph{Phys. Rev. D} {\bf
  90} (2014) 094503}, [\href{http://arxiv.org/abs/1407.6387}{{\tt 1407.6387}}].

\bibitem{Cadamuro:2010cz}
D.~Cadamuro, S.~Hannestad, G.~Raffelt and J.~Redondo, \emph{{Cosmological
  bounds on sub-MeV mass axions}},
  \href{http://dx.doi.org/10.1088/1475-7516/2011/02/003}{\emph{JCAP} {\bf 02}
  (2011) 003}, [\href{http://arxiv.org/abs/1011.3694}{{\tt 1011.3694}}].

\bibitem{Conlon:2013isa}
J.~P. Conlon and M.~C.~D. Marsh, \emph{{The Cosmophenomenology of Axionic Dark
  Radiation}}, \href{http://dx.doi.org/10.1007/JHEP10(2013)214}{\emph{JHEP}
  {\bf 10} (2013) 214}, [\href{http://arxiv.org/abs/1304.1804}{{\tt
  1304.1804}}].

\bibitem{Beutler:2011hx}
F.~Beutler, C.~Blake, M.~Colless, D.~H. Jones, L.~Staveley-Smith, L.~Campbell
  et~al., \emph{{The 6dF Galaxy Survey: Baryon Acoustic Oscillations and the
  Local Hubble Constant}},
  \href{http://dx.doi.org/10.1111/j.1365-2966.2011.19250.x}{\emph{Mon. Not.
  Roy. Astron. Soc.} {\bf 416} (2011) 3017--3032},
  [\href{http://arxiv.org/abs/1106.3366}{{\tt 1106.3366}}].

\bibitem{Ross:2014qpa}
A.~J. Ross, L.~Samushia, C.~Howlett, W.~J. Percival, A.~Burden and M.~Manera,
  \emph{{The clustering of the SDSS DR7 main Galaxy sample \textendash{} I. A 4
  per cent distance measure at $z = 0.15$}},
  \href{http://dx.doi.org/10.1093/mnras/stv154}{\emph{Mon. Not. Roy. Astron.
  Soc.} {\bf 449} (2015) 835--847}, [\href{http://arxiv.org/abs/1409.3242}{{\tt
  1409.3242}}].

\bibitem{Archidiacono:2019wdp}
M.~Archidiacono, D.~C. Hooper, R.~Murgia, S.~Bohr, J.~Lesgourgues and M.~Viel,
  \emph{{Constraining Dark Matter-Dark Radiation interactions with CMB, BAO,
  and Lyman-$\alpha$}},
  \href{http://dx.doi.org/10.1088/1475-7516/2019/10/055}{\emph{JCAP} {\bf 10}
  (2019) 055}, [\href{http://arxiv.org/abs/1907.01496}{{\tt 1907.01496}}].

\bibitem{Lewis:2002ah}
A.~Lewis and S.~Bridle, \emph{{Cosmological parameters from CMB and other data:
  A Monte Carlo approach}},
  \href{http://dx.doi.org/10.1103/PhysRevD.66.103511}{\emph{Phys. Rev. D} {\bf
  66} (2002) 103511}, [\href{http://arxiv.org/abs/astro-ph/0205436}{{\tt
  astro-ph/0205436}}].

\bibitem{Pisanti:2007hk}
O.~Pisanti, A.~Cirillo, S.~Esposito, F.~Iocco, G.~Mangano, G.~Miele et~al.,
  \emph{{PArthENoPE: Public Algorithm Evaluating the Nucleosynthesis of
  Primordial Elements}},
  \href{http://dx.doi.org/10.1016/j.cpc.2008.02.015}{\emph{Comput. Phys.
  Commun.} {\bf 178} (2008) 956--971},
  [\href{http://arxiv.org/abs/0705.0290}{{\tt 0705.0290}}].

\bibitem{Pitrou:2018cgg}
C.~Pitrou, A.~Coc, J.-P. Uzan and E.~Vangioni, \emph{{Precision big bang
  nucleosynthesis with improved Helium-4 predictions}},
  \href{http://dx.doi.org/10.1016/j.physrep.2018.04.005}{\emph{Phys. Rept.}
  {\bf 754} (2018) 1--66}, [\href{http://arxiv.org/abs/1801.08023}{{\tt
  1801.08023}}].

\bibitem{Lewis:2019xzd}
A.~Lewis, \emph{{GetDist: a Python package for analysing Monte Carlo samples}},
   \href{http://arxiv.org/abs/1910.13970}{{\tt 1910.13970}}.

\bibitem{DiValentino:2021izs}
E.~Di~Valentino, O.~Mena, S.~Pan, L.~Visinelli, W.~Yang, A.~Melchiorri et~al.,
  \emph{{In the realm of the Hubble tension\textemdash{}a review of
  solutions}}, \href{http://dx.doi.org/10.1088/1361-6382/ac086d}{\emph{Class.
  Quant. Grav.} {\bf 38} (2021) 153001},
  [\href{http://arxiv.org/abs/2103.01183}{{\tt 2103.01183}}].

\bibitem{CAST:2017uph}
{\scshape CAST} collaboration, V.~Anastassopoulos et~al., \emph{{New CAST Limit
  on the Axion-Photon Interaction}},
  \href{http://dx.doi.org/10.1038/nphys4109}{\emph{Nature Phys.} {\bf 13}
  (2017) 584--590}, [\href{http://arxiv.org/abs/1705.02290}{{\tt 1705.02290}}].

\bibitem{ciaran_o_hare_2020_3932430}
C.~O'Hare, \emph{cajohare/axionlimits: Axionlimits},  July, 2020.
\newblock 10.5281/zenodo.3932430.

\bibitem{Kim:1998va}
J.~E. Kim, \emph{{Constraints on very light axions from cavity experiments}},
  \href{http://dx.doi.org/10.1103/PhysRevD.58.055006}{\emph{Phys. Rev. D} {\bf
  58} (1998) 055006}, [\href{http://arxiv.org/abs/hep-ph/9802220}{{\tt
  hep-ph/9802220}}].

\bibitem{DiLuzio:2016sbl}
L.~Di~Luzio, F.~Mescia and E.~Nardi, \emph{{Redefining the Axion Window}},
  \href{http://dx.doi.org/10.1103/PhysRevLett.118.031801}{\emph{Phys. Rev.
  Lett.} {\bf 118} (2017) 031801}, [\href{http://arxiv.org/abs/1610.07593}{{\tt
  1610.07593}}].

\bibitem{DiLuzio:2017pfr}
L.~Di~Luzio, F.~Mescia and E.~Nardi, \emph{{Window for preferred axion
  models}}, \href{http://dx.doi.org/10.1103/PhysRevD.96.075003}{\emph{Phys.
  Rev. D} {\bf 96} (2017) 075003}, [\href{http://arxiv.org/abs/1705.05370}{{\tt
  1705.05370}}].

\bibitem{Dine:1981rt}
M.~Dine, W.~Fischler and M.~Srednicki, \emph{{A Simple Solution to the Strong
  CP Problem with a Harmless Axion}},
  \href{http://dx.doi.org/10.1016/0370-2693(81)90590-6}{\emph{Phys. Lett. B}
  {\bf 104} (1981) 199--202}.

\bibitem{Zhitnitsky:1980tq}
A.~R. Zhitnitsky, \emph{{On Possible Suppression of the Axion Hadron
  Interactions. (In Russian)}}, {\emph{Sov. J. Nucl. Phys.} {\bf 31} (1980)
  260}.

\bibitem{Kim:1979if}
J.~E. Kim, \emph{{Weak Interaction Singlet and Strong CP Invariance}},
  \href{http://dx.doi.org/10.1103/PhysRevLett.43.103}{\emph{Phys. Rev. Lett.}
  {\bf 43} (1979) 103}.

\bibitem{Shifman:1979if}
M.~A. Shifman, A.~I. Vainshtein and V.~I. Zakharov, \emph{{Can Confinement
  Ensure Natural CP Invariance of Strong Interactions?}},
  \href{http://dx.doi.org/10.1016/0550-3213(80)90209-6}{\emph{Nucl. Phys. B}
  {\bf 166} (1980) 493--506}.

\bibitem{Farina:2016tgd}
M.~Farina, D.~Pappadopulo, F.~Rompineve and A.~Tesi, \emph{{The photo-philic
  QCD axion}}, \href{http://dx.doi.org/10.1007/JHEP01(2017)095}{\emph{JHEP}
  {\bf 01} (2017) 095}, [\href{http://arxiv.org/abs/1611.09855}{{\tt
  1611.09855}}].

\bibitem{Agrawal:2017cmd}
P.~Agrawal, J.~Fan, M.~Reece and L.-T. Wang, \emph{{Experimental Targets for
  Photon Couplings of the QCD Axion}},
  \href{http://dx.doi.org/10.1007/JHEP02(2018)006}{\emph{JHEP} {\bf 02} (2018)
  006}, [\href{http://arxiv.org/abs/1709.06085}{{\tt 1709.06085}}].

\bibitem{Hook:2018jle}
A.~Hook, \emph{{Solving the Hierarchy Problem Discretely}},
  \href{http://dx.doi.org/10.1103/PhysRevLett.120.261802}{\emph{Phys. Rev.
  Lett.} {\bf 120} (2018) 261802}, [\href{http://arxiv.org/abs/1802.10093}{{\tt
  1802.10093}}].

\bibitem{DiLuzio:2021pxd}
L.~Di~Luzio, B.~Gavela, P.~Quilez and A.~Ringwald, \emph{{An even lighter QCD
  axion}}, \href{http://dx.doi.org/10.1007/JHEP05(2021)184}{\emph{JHEP} {\bf
  05} (2021) 184}, [\href{http://arxiv.org/abs/2102.00012}{{\tt 2102.00012}}].

\bibitem{Sokolov:2021ydn}
A.~V. Sokolov and A.~Ringwald, \emph{{Photophilic hadronic axion from heavy
  magnetic monopoles}},
  \href{http://dx.doi.org/10.1007/JHEP06(2021)123}{\emph{JHEP} {\bf 06} (2021)
  123}, [\href{http://arxiv.org/abs/2104.02574}{{\tt 2104.02574}}].

\bibitem{ADMX:2021nhd}
{\scshape ADMX} collaboration, C.~Bartram et~al., \emph{{Search for Invisible
  Axion Dark Matter in the 3.3\textendash{}4.2\,\,\ensuremath{\mu}eV Mass
  Range}}, \href{http://dx.doi.org/10.1103/PhysRevLett.127.261803}{\emph{Phys.
  Rev. Lett.} {\bf 127} (2021) 261803},
  [\href{http://arxiv.org/abs/2110.06096}{{\tt 2110.06096}}].

\bibitem{Friedland:2012hj}
A.~Friedland, M.~Giannotti and M.~Wise, \emph{{Constraining the Axion-Photon
  Coupling with Massive Stars}},
  \href{http://dx.doi.org/10.1103/PhysRevLett.110.061101}{\emph{Phys. Rev.
  Lett.} {\bf 110} (2013) 061101}, [\href{http://arxiv.org/abs/1210.1271}{{\tt
  1210.1271}}].

\bibitem{Ayala:2014pea}
A.~Ayala, I.~Dom\'\i{}nguez, M.~Giannotti, A.~Mirizzi and O.~Straniero,
  \emph{{Revisiting the bound on axion-photon coupling from Globular
  Clusters}},
  \href{http://dx.doi.org/10.1103/PhysRevLett.113.191302}{\emph{Phys. Rev.
  Lett.} {\bf 113} (2014) 191302}, [\href{http://arxiv.org/abs/1406.6053}{{\tt
  1406.6053}}].

\bibitem{Lucente:2022vuo}
G.~Lucente, L.~Mastrototaro, P.~Carenza, L.~Di~Luzio, M.~Giannotti and
  A.~Mirizzi, \emph{{Axion signatures from supernova explosions through the
  nucleon electric-dipole portal}},
  \href{http://dx.doi.org/10.1103/PhysRevD.105.123020}{\emph{Phys. Rev. D} {\bf
  105} (2022) 123020}, [\href{http://arxiv.org/abs/2203.15812}{{\tt
  2203.15812}}].

\bibitem{DEramo:2022nvb}
F.~D'Eramo, E.~Di~Valentino, W.~Giar\`e, F.~Hajkarim, A.~Melchiorri, O.~Mena
  et~al., \emph{{Cosmological Bound on the QCD Axion Mass, Redux}},
  \href{http://arxiv.org/abs/2205.07849}{{\tt 2205.07849}}.

\end{thebibliography}\endgroup

\end{document}